\documentclass[fleqn,usenatbib]{mnras}

\usepackage{newtxtext,newtxmath}

\usepackage[T1]{fontenc}

\usepackage{graphicx}
\usepackage{subcaption}  
\usepackage{physics} 
\usepackage{booktabs} 
\usepackage[flushleft]{threeparttable} 
\usepackage{cancel} 
\usepackage{bm}
\usepackage[svgnames]{xcolor}    
\usepackage{ulem}

\usepackage[dvipsnames]{xcolor} 

\DeclareRobustCommand{\VAN}[3]{#2}
\let\VANthebibliography\thebibliography
\def\thebibliography{\DeclareRobustCommand{\VAN}[3]{##3}\VANthebibliography}

\usepackage{graphicx}	
\usepackage{amsmath}	
\usepackage{verbatim}

\title[Bar-halo interaction]{Bar-halo interaction: the role of orbital anisotropy}

\author[P. Gherghinescu et al.]{
Paula Gherghinescu$^{1}$,\thanks{E-mail: paula.gherghinescu@durham.ac.uk }
Francesca Fragkoudi$^{1,2}$,
Alis Deason$^{1,3}$ 
\\
$^{1}$Institute for Computational Cosmology, Department of Physics, University of Durham, South Road, Durham DH1 3LE, UK\\
$^{2}$School of Sciences, European University Cyprus, Diogenes Street, Engomi, 1516 Nicosia, Cyprus \\
$^{3}$Centre for Extragalactic Astronomy, Department of Physics, University of Durham, South Road, Durham DH1 3LE, UK\\
}

\date{Accepted XXX. Received YYY; in original form ZZZ}

\pubyear{\the\year{}}

\begin{document}
\label{firstpage}
\pagerange{\pageref{firstpage}--\pageref{lastpage}}
\maketitle

\begin{abstract}
We investigate the dynamical response of dispersion-dominated halo populations to a rotating galactic bar, focusing on how the underlying halo phase space distribution function (DF), and in particular the orbital anisotropy, shapes resonant structure formation. Using controlled test-particle simulations in a fixed Milky Way–like potential, we systematically vary the velocity anisotropy and net rotation of halo-like components while keeping the halo density profile, global potential, and bar properties fixed.
We find that bar-induced resonances generate prominent substructure in energy–angular momentum space, but that the morphology, strength, and density contrast (i.e. overdensities versus underdensities) of these features depend sensitively on the halo orbital anisotropy and how resonant transport aligns with gradients of the DF in action space. For instance, radially biased halos tend to exhibit stronger responses and features across all main resonances. Our results also show that angular momentum exchange and the torque exerted on a halo are governed not only by its density profile but crucially by its orbital anisotropy structure. This highlights the importance of halo anisotropy when interpreting phase-space substructure in the stellar halo of the Milky Way with current and future surveys, while also having implications in further understanding the DM halo-bar coupling in disk galaxies.
\end{abstract}

\begin{keywords}
galaxies: kinematics and dynamics -- galaxies: haloes
\end{keywords}



\section{Introduction} \label{sec:intro}
Dispersion-dominated stellar and dark matter systems, ranging from galactic bulges and stellar halos to dark matter (DM) halos, exhibit broad orbital distributions, with stars and particles moving on radial, isotropic, or retrograde trajectories across a wide range of energies. This diversity means that their dynamical response to perturbations can differ significantly from that of rotationally supported disks.

\par Many disk galaxies, including our own Milky Way (hereafter MW), host a prominent stellar bar: a strong, rotating non-axisymmetric structure that strongly alters the gravitational potential of the inner galaxy. Bars redistribute angular momentum through resonant torques and trapping, reshaping orbital families and driving secular evolution \citep[e.g.][]{LBK_1972,Weinberg1985,Debattista_Sellwood2000,Athanassoula2003,Fragkoudi+2015,Fragkoudi+2016}. In disks, this process gives rise to prominent dynamical features such as phase space ridges, moving groups, and velocity substructures \citep[e.g.][]{Dehnen+2000,Fux+2001,Anatoja2018,Kawata+2018,Hunt+2019, Fragkoudi2019}.

\par Despite the extensive study of bar-disk interactions, the interaction between bars and dispersion-dominated populations, especially stellar halos, remains understudied. The interaction between bars and DM halos has been explored more in both analytic theory and numerical simulations. Resonant interactions can transfer angular momentum between the bar and halo, leading to bar slow-down, inner halo shape changes such as “shadow bars” or wakes, and reorientation of halo orbits in both prograde and retrograde halos \citep[e.g.][]{Tremain_Weinberg1984,Debattista_Sellwood2000,Athanassoula2003,Weinberg_Katz2007,Petersen+2016_shadow_bar,Collier_Madigan2021,Fragkoudi+2021}. 

\par Recently, the interaction between the Milky Way's stellar halo and the bar has started to receive increasing attention. While much of the classic work focused on DM halos, several studies have begun to explore how stellar halo stars respond to the bar, suggesting that stellar halos may exhibit bar-induced structures \citep[e.g.,][]{Dilamore+2023,Dillamore+2024,Tomlinson2026}. This growing body of research highlights the need to systematically understand how the bar interacts with dispersion-dominated stellar populations, motivating the present study.

\par In this work, we investigate how the underlying distribution function (DF) of a dispersion-dominated halo-like component shapes its dynamical response to a barred potential. For stellar halos, a large orbital diversity is expected and reflects the complex assembly history of the halo, including accretion and merger events, and implies that the present-day DF is shaped by its formation pathway \citep[e.g.,][]{Bullock_Johnston2005,Cooper+2010}. For DM halos, they are typically less radially anisotropic compared to stellar halos, but are likewise characterised by non-isotropic velocity distributions and a range of velocity anisotropy profiles \citep[e.g.,][]{He+2024,Zhang+2026}. We therefore expect that different halo DFs will produce correspondingly different dynamical responses when interacting with a large, time-dependent perturbation such as the galactic bar. 

\par To investigate this, we employ controlled test-particle simulations in a fixed MW-like potential with a time-dependent bar component. In all experiments, the global potential and bar properties are held fixed, so that any differences in the halo response arise solely from variations in the initial DF.

\par We consider haloes that share the same density profile but differ in their orbital structure, spanning a range of velocity anisotropies from radially to tangentially biased distributions, as well as varying degrees of net rotation. Our goal is to systematically characterise how the dynamical response of a halo depends on its underlying orbital composition.

\par In Section \ref{sec:method}, we introduce the test-particle simulation setup and describe the properties of the halo models considered. In Section \ref{sec:results}, we present the resultant dynamical response of these halos to the bar-like perturbation, analysing the emergent structures in energy–angular momentum and action space, highlighting how these depend on the underlying DF. Finally, in Section \ref{sec:discussion}, we interpret our findings in the context of perturbation theory and resonant dynamics, with a particular focus on how these mechanisms operate in dispersion-dominated systems.  While our work has natural applications to stellar halos, the findings are equally relevant to DM halos as well. Finally, in Section \ref{sec:conclusion} we present our concluding remarks.

\section{Method} \label{sec:method}
We employ a suite of test-particle simulations, with different halo DFs, to investigate the dynamical interaction between a rotating Galactic bar and halo-like orbits. The halo is modelled as a collisionless tracer population that does not contribute to the total Galactic potential; in particular, self-gravity and back-reaction on the potential are neglected. The particles are evolved in a fixed, Milky Way–like potential that includes a rotating bar with properties representative of the Milky Way. This setup isolates the orbital response of the test particles to the time-dependent, non-axisymmetric perturbation introduced by the bar, enabling a controlled study of bar–halo interactions.

\subsection{Milky Way-like gravitational potential}
We adopt the Milky Way potential implemented in Agama \citep{Vasiliev2019_agama}, following the barred model described in \cite{Hunter+2024}. This implementation incorporates the analytic bar model of \cite{Sormani+2022}, which is an analytic approximation of the bar model from \cite{Portail+2017}, and includes axisymmetric components such as thin and thick stellar disks, a nuclear stellar disk, gas disks, and a dark matter halo. The Agama implementation of this potential is publicly available\footnote{\url{https://github.com/GalacticDynamics-Oxford/Agama}}. The bar is assigned a fixed pattern speed of $\Omega_{b} = 37.5\ \mathrm{km;s^{-1};kpc^{-1}}$ and the present-day orientation angle of $\phi_b = 25^\circ$.

\subsection{Test particles and initial conditions}
To construct halo models with controlled orbital structure, we make use of action-based distribution functions. Action-angle coordinates $(\mathbf{J},\boldsymbol{\theta})$ are widely used in galactic dynamics, where the actions $\mathbf{J}=(J_r,J_\phi,J_z)$ provide conserved orbital labels associated with the radial, azimuthal, and vertical motion of an orbit: $J_r$ quantifies the extent of radial motion, $J_z$ measures the vertical excursion from the plane, and $J_\phi$ corresponds to the azimuthal angular momentum \citep[e.g.,][]{Binney_Tremaine_2008}. 

\par Distribution functions expressed in terms of actions therefore provide a convenient framework for constructing equilibrium dynamical models with tunable velocity anisotropy and net rotation. We draw halo test particles from an action-based double power-law DF, suitable for describing spheroidal systems, as presented in \cite{Posti+2015} and adapted and implemented in Agama:
\begin{align} 
\begin{split}
   & f(\vb*{J})  = \frac{M_{0}}{(2\pi J_{0})^{3}}\;
   \biggl[ 1 + \left( \frac{J_{0}}{h(\vb*{J})} \right)^\eta \biggr]^{\alpha/\eta}\;
   \biggl[ 1 + \left( \frac{g(\vb*{J})}{J_{0}} \right)^\eta \biggr]^{-\beta/\eta} \\
   &\mbox{}\qquad\times  \biggl(1+\chi \tanh\frac{J_{\phi}}{J_{\phi,0}}\biggr), \\ 
& \text{where}\,\,g(\vb*{J}) = g_{r}J_{r} + g_{z}J_{z} + (3- g_{r}-g_{z})\left | J_{\phi} \right |, \\
& \text{and}\,\;\;\;\,h(\vb*{J}) = h_{r}J_{r} + h_{z}J_{z} + (3- h_{r}-h_{z})\left | J_{\phi} \right |.
\end{split}
\label{eq:double_power_law_df}
\end{align}

Here, $\alpha$ and $\beta$ are power-law indices that control the density profile in the inner and outer regions of the halo, respectively. The functions $h(\mathbf{J})$ and $g(\mathbf{J})$ are linear combinations of actions with mixing coefficients ${h_r,h_z,h_\phi}$ and ${g_r,g_z,g_\phi}$, each summing to 3, leaving two independent parameters per combination. These coefficients set the flattening and velocity ellipsoids of the halo tracers in the inner and outer regions. Finally, the term $(1+\chi \tanh(J_\phi/J_{\phi,0}))$ introduces a net rotation, controlled by $\chi$, around the $z$-axis.

\par The DF depends on the axisymmetric actions and allows explicit control over the orbital anisotropy profile,
\begin{equation}
\beta(r) = 1 - \frac{\sigma_\theta^2 + \sigma_\phi^2}{2 \sigma_r^2}.
\end{equation}
Here, ($\sigma_r,\sigma_\theta,\sigma_\phi$) are the velocity dispersions in spherical polar coordinates. Therefore, it can be seen that $\beta(r)$ characterizes the relative contribution of radial and tangential motions:
\begin{itemize}
\item Radially biased halos ($\beta>0$) are dominated by eccentric, plunging orbits that penetrate the inner Galaxy.
\item Tangentially biased halos ($\beta<0$) preferentially populate high-angular-momentum and more circular orbits.
\end{itemize}
Cosmological simulations indicate that stellar haloes generally exhibit radially increasing anisotropy, with orbits becoming progressively more radial at larger galactocentric radii \citep[e.g.,][]{Amorisco2017}. DM haloes likewise tend to exhibit radially anisotropic velocity distributions, although typically with lower levels of anisotropy than their stellar halo counterparts \citep[e.g.,][]{Cole_Lacey1996,Hansen_Moore2006,He+2024}.

To explore how orbital structure influences the dynamical response to the bar, we construct a suite of four halo DFs that share a common density profile but differ in velocity anisotropy and net rotation. Specifically, we consider a baseline radially anisotropic halo with prograde rotation (ProOG), and vary the orbital structure in two ways (i) the degree of velocity anisotropy, ranging from strongly radial to tangentially biased systems, and (ii) the sense of net rotation, aligned or counter-aligned with the bar.

\par
\begin{figure}
    \includegraphics[width=0.5\textwidth]{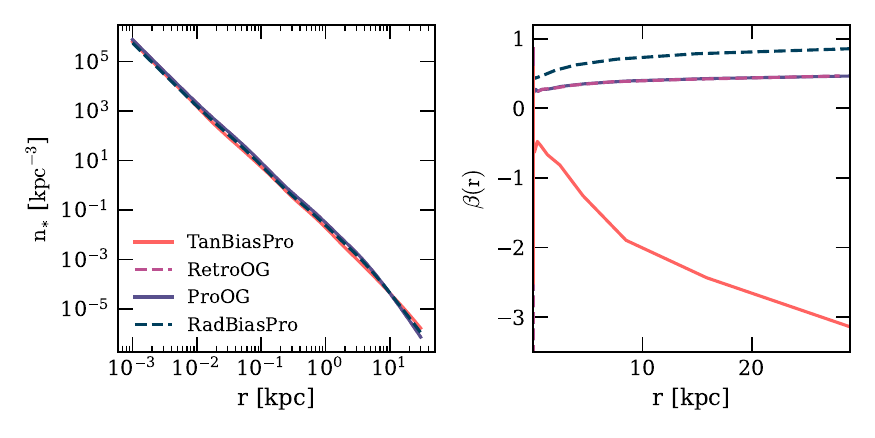}
    \caption{Density and ansitropoy profiles corresponding to all the DFs used to generate ICs.}
    \label{fig:ndens_beta_allsims}
\end{figure}

 We summarise the test particle simulation properties in Table \ref{tab:df_parameters} and below. In Figure \ref{fig:ndens_beta_allsims} it can be seen that the tracer halos have the same density distribution, but differ in their orbital anisotropy profiles.

\begin{itemize}

\item \textbf{ProgradeOG}: The base simulation features a moderately radially anisotropic halo and a net \textbf{prograde} rotation aligned with the bar.

\item \textbf{RetrogradeOG}: Identical to Prograde OG in its anisotropy and density profile, but with a net \textbf{retrograde} rotation (i.e. opposite to the bar’s rotation).

\item \textbf{RadBiasPro}: A \textbf{strongly radially anisotropic} halo with a high degree of radial anisotropy, combined with \textbf{prograde} net rotation.

\item \textbf{TanBiasPro}: A halo with \textbf{predominantly tangentially} biased orbits, also with \textbf{prograde} rotation aligned with the bar.
\end{itemize}

To generate the initial conditions (ICs) for all four simulations, we sample $N_* = 10^7$ positions and velocities ($\bm{x,v}$) = ($x,y,z,v_x,v_y,v_z$) from the axisymmetric DFs. The particles are then numerically integrated in the full time-dependent potential using the orbit-integration routines implemented in Agama. The bar amplitude is grown adiabatically from zero to its maximum value over $1\,\mathrm{Gyr}$, after which the particles are evolved for a further $4\,\mathrm{Gyr}$ in the fully developed barred potential.

\begin{table*}
  \caption{\label{tab:df_parameters} The parameters values of DFs used to generate the ICs for all the simulations.}
  \centering 
  \begin{threeparttable}
    \begin{tabular*}{0.8\linewidth}{@{\extracolsep{\fill}}ccccc}
\midrule
    DF parameters  & ProOG & RetroOG & RadBiasPro & TanBiasPro \\
     \midrule
 $ \begin{aligned}
    \alpha \\
    \beta \\
    \eta \\
    J_{0} \: [\mathrm{kpc\,km\,s^{-1}}]\\
    h_{r} \\
    h_{z} \\
    g_{r} \\
    g_{z} \\
    \chi \\
    J_{\phi,0} \: [ \mathrm{kpc\,km\,s^{-1}}]
    \end{aligned} $ &
    $\begin{aligned} 
    & 2.5\\
    & 5.5 \\
    & 1 \\
    & 8000\\
    & 0.75 \\
    & 1.7 \\
    & 0.88 \\
    & 1.1 \\
    & -0.3 \\
    & 0 \\
    
    \end{aligned}$  
&
    $\begin{aligned} 
    & 2.5\\
    & 5.5 \\
    & 1 \\
    & 8000 \\
    & 0.75 \\
    & 1.7 \\
    & 0.88 \\
    & 1.1 \\
    & +0.3 \\
    & 0  \\
    
    \end{aligned}$  

&
    $\begin{aligned} 
    & 2.5\\
    & 5.5 \\
    & 1 \\
    & 8000 \\
    & 0.5 \\
    & 1.3 \\
    & 0.1 \\
    & 1.1 \\
    & -0.3 \\
    & 0  \\
    
    \end{aligned}$  

&
    $\begin{aligned} 
    & 2.5\\
    & 5.5 \\
    & 1 \\
    & 8000 \\
    & 2 \\
    & 0.2 \\
    & 2.2 \\
    & 0.1 \\
    & -0.3 \\
    & 0  \\
    
    \end{aligned}$  

\\ 
    \midrule
    \end{tabular*}
\end{threeparttable}

\end{table*}

\section{Results} \label{sec:results}

In this section, we present the dynamical response of halo tracer particles to the growth of a rotating Galactic bar. We focus on the emergence of substructure in E-L$_z$ space and link these features directly to the underlying orbital dynamics. We first examine the development of non-axisymmetric structures in the E–L$_z$ plane across all simulations (Sec. \ref{subsec:ELz_substruct}), and relate their appearance to the presence of bar-induced resonances. We then analyse the corresponding structures in action space (Sec. \ref{subsec:ELz-actions}), in order to further connect the E–L$_z$ substructures to the orbital composition of the halo.

\par To identify resonant halo particles, we estimate the fundamental orbital frequencies using a Fourier-based spectral analysis. For each particle, we record the time series of its phase-space coordinates over a fixed interval of $\Delta t = 2$ Gyr in the fully developed barred potential, and extract the dominant radial, azimuthal, and vertical frequencies $(\Omega_r,\Omega_{\phi},\Omega_{z})$ via a fast Fourier transform (FFT). Orbital frequencies computed this way, remain a robust diagnostic of orbital structure at resonances and provide effective orbit labels. Resonant particles are identified using the frequency resonance condition:
\begin{equation}
    l\Omega_r + m\Omega_{\phi} \approx m\Omega_{bar},
    \label{eq:resonance_criteria}
\end{equation}
where $(l,m)$ are integers and $\Omega_{bar}$ is the bar pattern speed. For example, some of the main resonances investigated in this work  are corotation $(l,m=0,2)$, 1:1 $(l/m=1/1)$, Inner Lindblad resonance (ILR) $(l/m=-1/2)$, the 1:4 ultraharmonic $(l/m=-1/4)$, as well as the higher order $(l/m=3/2)$.

\subsection{E-$\mathrm{L_{z}}$ substructures} \label{subsec:ELz_substruct}
\par
\begin{figure*}
        \centering
        \includegraphics[width=\linewidth]{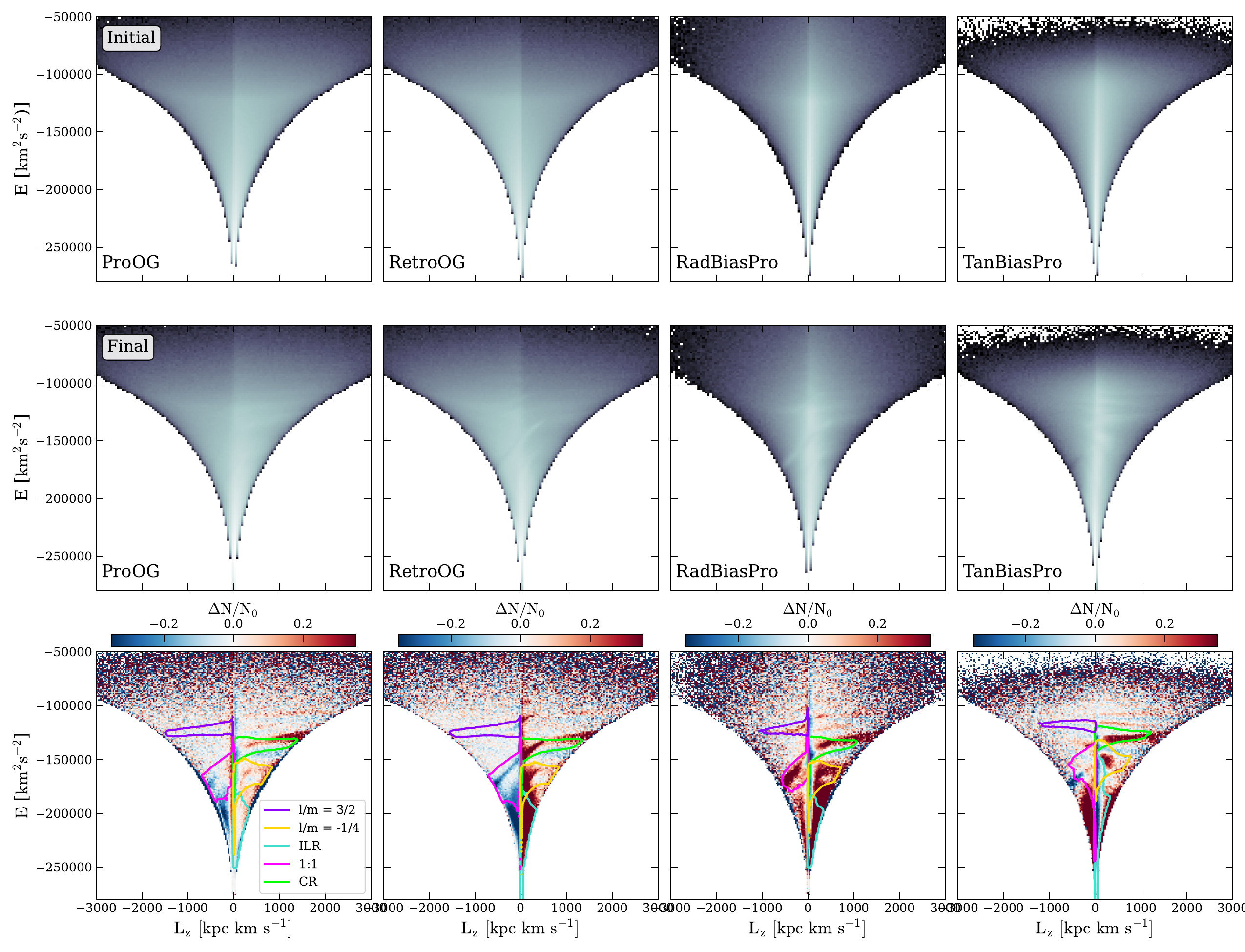}
        \caption{E-Lz all simulations (a) first row: initial conditions, (b) middle row: final snapshot, (c) bottom row: density contrast between final and initial to highlight emergent substructures. All stars are within $r\leq 12$ kpc from the galaxy's center.}
        \label{fig:ELz_summary_allsims}

\end{figure*}

 In Fig. \ref{fig:ELz_summary_allsims}, we summarise the E–L$_z$ diagrams for all simulations. The first row shows the initial conditions prior to the introduction of the bar. As expected, the distribution is axisymmetric and featureless, with no ridges or substructures present. We can clearly see that the ProgradeOG and RetrogradeOG models show a clear asymmetry in $L_z$, with enhanced density at positive and negative $L_z$, respectively, reflecting their net rotation. It can also be seen that radially biased model exhibits a higher concentration of particles near $L_z \approx 0$, consistent with its predominance of low angular momentum orbits.

\par The second row of Fig. \ref{fig:ELz_summary_allsims} shows the E–L$_z$ diagrams at the final snapshot, after the bar has been fully grown and the system has been evolved. Non-axisymmetric substructures clearly emerge, reflecting the dynamical response of the halo particles to the time-dependent bar potential.

\par The third row of Fig. \ref{fig:ELz_summary_allsims} presents the density contrast per E–L$_z$ bin between the final and initial snapshots, highlighting regions that have become overdense (red) or underdense (blue) compared to the initial distribution. Coloured contours indicate the locations of particles trapped in resonances (according to eq. \ref{eq:resonance_criteria}), showing a clear correspondence between the emergent density features and resonant regions. The largest changes occur at the 1:1 retrograde (pink), corotation (green), ILR (cyan) and $l/m=-1/4$ ultraharmonic (yellow) resonances.
\par Overall, the final E–L$_z$ structures vary across simulations. Some simulations exhibit overdensities where others show underdensities at the same resonance. For example, the retrograde 1:1 resonance typically produces underdensities of varying strength across most simulations, except in the radially biased case (RadBiasPro), where a strong overdensity is found. On the other hand, it be seen that on the prograde side, the corotation resonance is associated with a strong overdensity, accompanied by a weaker underdensity at slightly lower energies for all four simulations. The E-L$_z$ location of these resonances is, however, the same, which is a consequence of the fixed bar and global potential.

Since the global potential and bar properties are identical in all experiments, and no additional processes are present (e.g. mergers, interactions, or self-gravity), these differences indicate that the underlying DF of the tracers plays the dominant role in shaping the emergent substructure. In particular, the sign and strength of resonant features depend on the gradients of the DF along the direction of resonant transport in phase space: resonances can produce either over- or under-densities depending on whether particles are redistributed from regions of lower to higher phase-space density, or vice versa. We dive into this further and discuss the phenomenology in detail in Section \ref{sec:discussion}.

\subsection{Orbital redistribution in E–L$_z$ space: insights from actions} \label{subsec:ELz-actions}
\begin{figure*}
    \centering
    
    \begin{subfigure}{\linewidth}
        \centering
        \includegraphics[width=\linewidth]{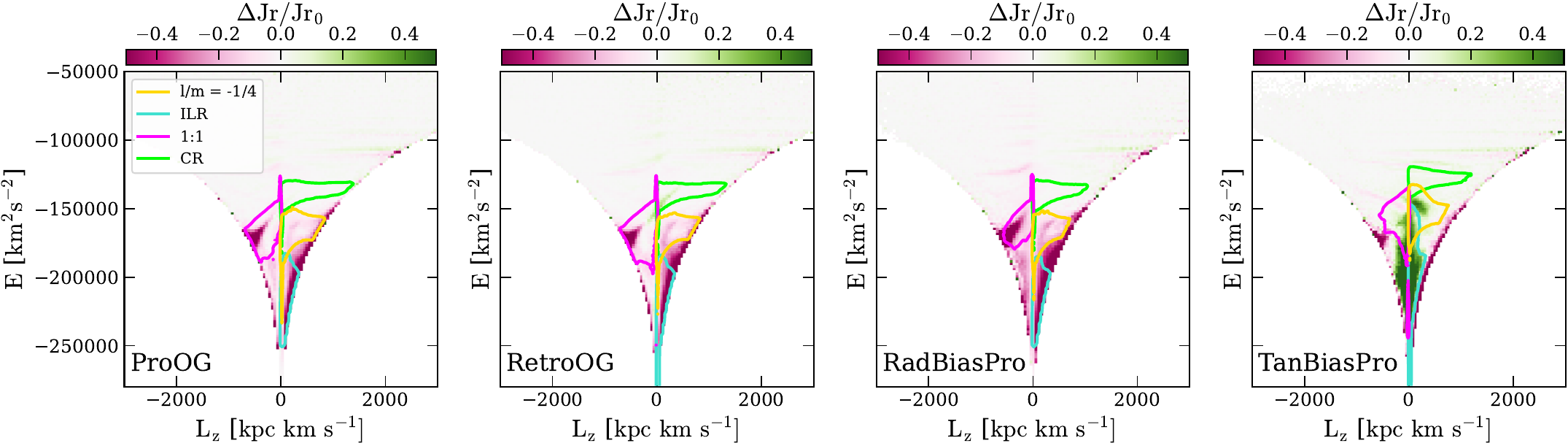}
    \end{subfigure}
    \hfill
    \begin{subfigure}{\linewidth}
        \centering
        \includegraphics[width=\linewidth]{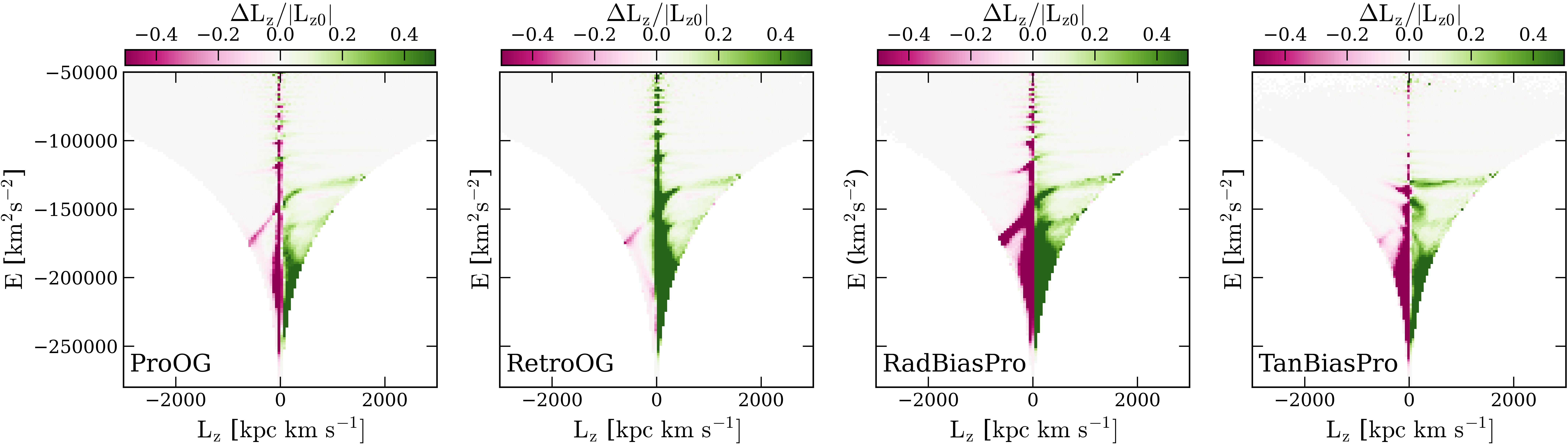}
    \end{subfigure}

    \hfill
    \begin{subfigure}{\linewidth}
        \centering
        \includegraphics[width=\linewidth]{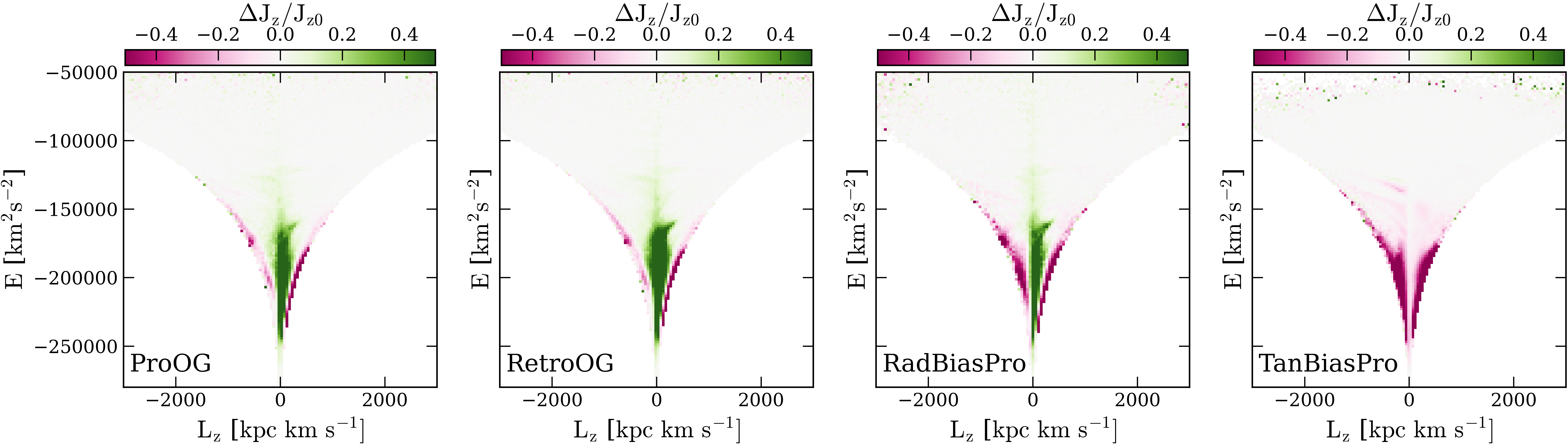}
    \end{subfigure}

    \hfill
    \begin{subfigure}{\linewidth}
        \centering
        \includegraphics[width=\linewidth]{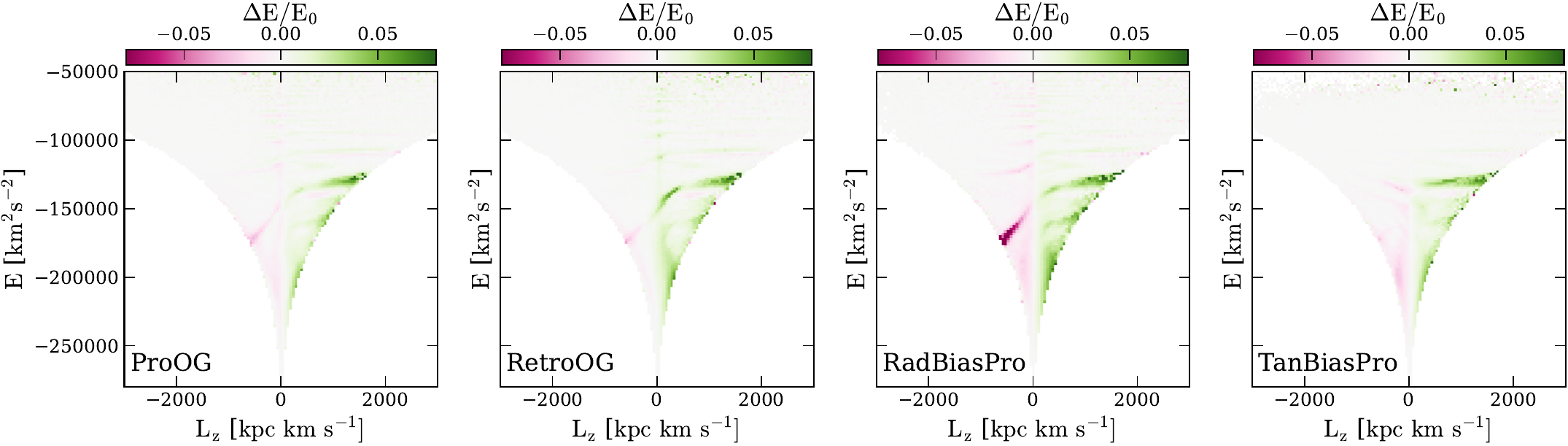}
        \label{fig:ELz_final_cbar_delta_E}
    \end{subfigure}

    \caption{Energy - angular momentum plots at final snapshot, for particles within $r\leq 12 $ kpc, with colorbar indicating fractional changes in the (a) radial action $J_r$, (b) azimuthal angular momentum $L_z$, (c) vertical action $J_z$, and (d) total orbital energy $E$.}
    \label{fig:ELz_final_cbar_delta_actions}
    
\end{figure*}

To investigate how the bar redistributes halo particles in orbital space, we examine changes in the actions of particles projected onto the E–L$z$ plane. We point out that standard methods of action computation (eg., St\"ackel fudge) assume regular, non-resonant orbits, and this assumption breaks down for resonant trajectories, leading to increased uncertainties and occasional inaccuracies, particularly in the radial action $J_{r}$. Despite this limitation, the computed actions remain useful as approximate orbit labels, capturing the relative ordering and character of orbits \citep[e.g.,][]{Trick+2021}. To prevent computational failures from contaminating our analysis, we remove particles with very high action values of the order $>\mathcal{O}(10^4)$ from the datasets.
The resulting E–L$_z$ maps presented in Fig. \ref{fig:ELz_final_cbar_delta_actions}, color-coded by changes in actions, provide a direct visualization of how orbits at different resonances are redistributed by the bar.

\par Fractional changes in the radial action $J_{r}$ are shown in the top row of Fig. ~\ref{fig:ELz_final_cbar_delta_actions}, those in the azimuthal angular momentum $L_{z}$ in the second row, and finally changes in the vertical action $J_{z}$ in the third row Fig.~\ref{fig:ELz_final_cbar_delta_actions}. Significant changes occur primarily at resonances, as the bar perturbation redistributes resonant particles in orbital space. 

\par At resonances, particles are redistributed in E-L$_z$ space while approximately conserving the Jacobi energy, $E_J = E-\Omega_{\rm bar}L_z$, and therefore satisfy:
\begin{equation}
    \Delta E = \Omega_{\rm bar}\Delta L_z.
    \label{eq:EJacobi_deltaE_deltaLz}
\end{equation}

Similarly, in action space, resonant transport proceeds along directions set by the resonance according to \citep[e.g.,][]{Sellwood+2002}:
\begin{equation} 
\Delta J_r / \Delta L_z = l/m,
\label{eq:deltaJr_deltaLz}
\end{equation}
where $(l,m)$ specify the resonance (see Sec. \ref{sec:discussion} for derivation and further discussion). The changes in radial ($J_r$) and azimuthal ($J_{\phi}=L_z$) actions shown in the first and second rows of Fig.~\ref{fig:ELz_final_cbar_delta_actions} follow these expected trends. For example, at corotation ($l/m=0$), stars gain angular momentum while experiencing little change in radial action ($\Delta J_r\approx0$), indicating angular momentum exchange with minimal orbital heating. We expand more on this in the next Section \ref{sec:discussion}, where we dive in detail into the mechanisms that cause angular momentum gain or loss at resonances as well as how that relates to the underlying DF.

\par The bottom panel of Fig.~\ref{fig:ELz_final_cbar_delta_actions} shows the corresponding changes in total orbital energy. Energy evolution is likewise concentrated at resonances. On average, resonant stars on the prograde side tend to gain angular momentum ($\Delta L_z>0$), and therefore, according to eq.~\ref{eq:EJacobi_deltaE_deltaLz}, also gain energy ($\Delta E>0$). Conversely, resonant stars on the retrograde side typically lose angular momentum and correspondingly decrease their orbital energy. These trends and their physical motivation are explained in detail in the next Section \ref{sec:discussion}.

\par In contrast, the vertical action $J_z$ exhibits less pronounced evolution than $J_r$ and $L_z$ at the main resonances. Nevertheless, increases in $J_z$ are visible for low-$L_z$ stars and the lower $L_{z}$ region of the $l/m=-1/4$ ultraharmonic for three of the simulations: ProOG, RetroOG, and RadBiasPro. Furthermore, across all simulations it can be seen that highest L$_z$ stars (at each energy $E$) appear to reduce their vertical action $J_z$ on both the prograde and retrograde side.

\par \par In the next Section \ref{sec:discussion}, we interpret these results within the theoretical resonant dynamics framework, and relate the observed phase-space structures and orbital redistribution to the underlying halo DF.

\subsection{Metallicities}

From our results, it is clear that we expect the bar to redistribute stars in the stellar halo. A possible way of tracing this redistribution could be through metallicities $\mathrm{[Fe/H]}$. To illustrate this point, we assign a radial metallicity gradient to the test particles ICs to connect the dynamical redistribution induced by the bar to observable signatures as follows:

\begin{equation}
    [Fe/H]_{i}=-10^{-2}a_{i}-1.5,
\end{equation}
with $a_i$ the apocenter of particle $i$ at the initial snapshot, before the bar perturbation is introduced. The apocenters are computed as the apocenter radii corresponding to a planar orbit with given energy $E_i$ and angular momentum $L_{z,i}$.

The top row of Fig. ~\ref{fig:ELz_final_cbar_delta_actions} shows the distribution of stars in E–L$_z$ at the final snapshot for all the stars within $r\leq12$ kpc, with the colour scale indicating the mean metallicity in each bin. The second row of Fig. \ref{fig:ELz_final_cbar_delta_actions} shows our “solar neighbourhood” selection, defined as a spherical region of radius $r{\rm_{sph}}=4$ kpc centred at $(x,y,z)=(-8,0,0)$ kpc, and $ \approx -25^{\circ}$ behind the bar. However, we emphasise that this region is not intended to represent a tailored Milky Way analogue. While the underlying potential is MW–like, the simulation is not calibrated to reproduce chemo-dynamical structure of the MW's halo. The selection should therefore be interpreted as a qualitative solar-neighbourhood–like region.

\par The emergent metallicity substructures are highlighted in the third and fourth rows of Fig. \ref{fig:ELz_final_cbar_delta_actions}, where we show the same $E-L_z$ plots as before, but with the colorbar indicating the \textit{metallicity change} $\Delta [Fe/H]$ per bin between initial and final snapshots. It can be seen, that on average on the prograde side, the identified resonances drive localized metal enrichment, producing metal-rich substructures, with the strongest effect seen at corotation. Conversely, on the retrograde side, resonances predominantly redistribute stars into metal-poor substructures, with the retrograde 1:1 resonance producing a clear metal poor signature (compared to the 'background' $\mathrm{[Fe/H]}$ distribution), but also higher order resonances as well (eg, $l/m=3/2$). These trends are in line with expectations drawn from the previous subsection, as we can clearly see that on average the retrograde resonances increase $E$ and $Lz$, transporting stars from the inner, more metal rich regions of the halo, while the converse is true for the retrograde side. It is to be noted that, for the RetroOG simulation near $L_z \simeq 0$, the 1:1 resonance produces localized metal enrichment, in line with the $E-Lz$ transport identified in Section \ref{subsec:ELz_substruct}. 

\par We do stress that these metallicity signatures cannot be generalised, since the resulting patterns depend on the assumed pre-existing metallicity distribution, which will vary from halo to halo. Rather, this represents a controlled and illustrative experiment demonstrating that metallicities may provide a useful tracer of stars that have been redistributed by the bar.

\begin{figure*}
    \centering
    \begin{subfigure}{\linewidth}
        \centering
        \includegraphics[width=\linewidth]{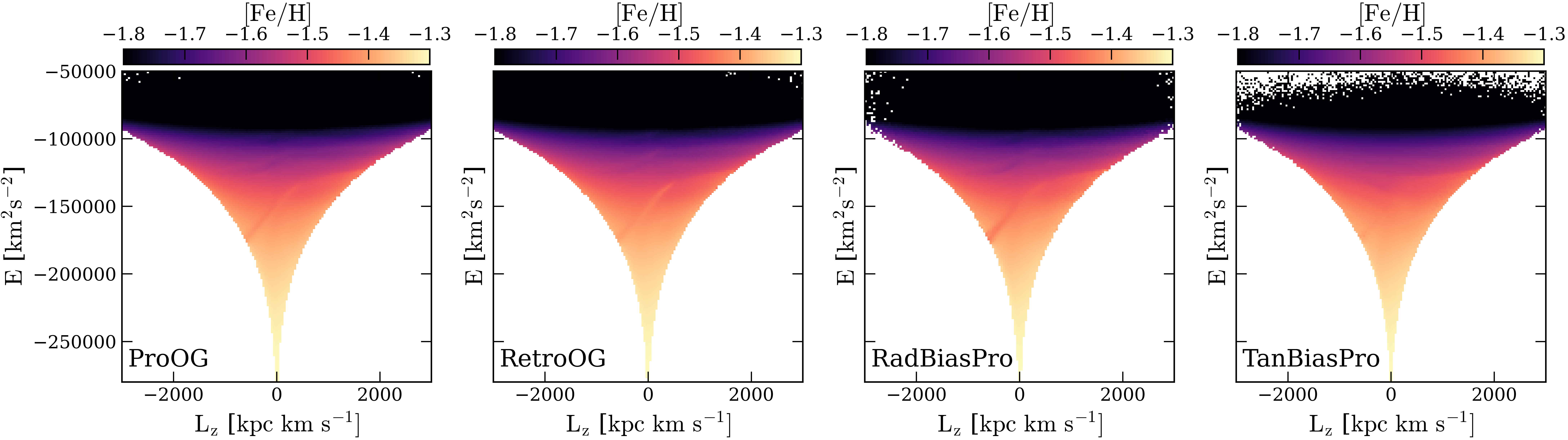}
        \label{fig:ELz_feh}
    \end{subfigure}
    \hfill
    \begin{subfigure}{\linewidth}
        \centering
        \includegraphics[width=\linewidth]{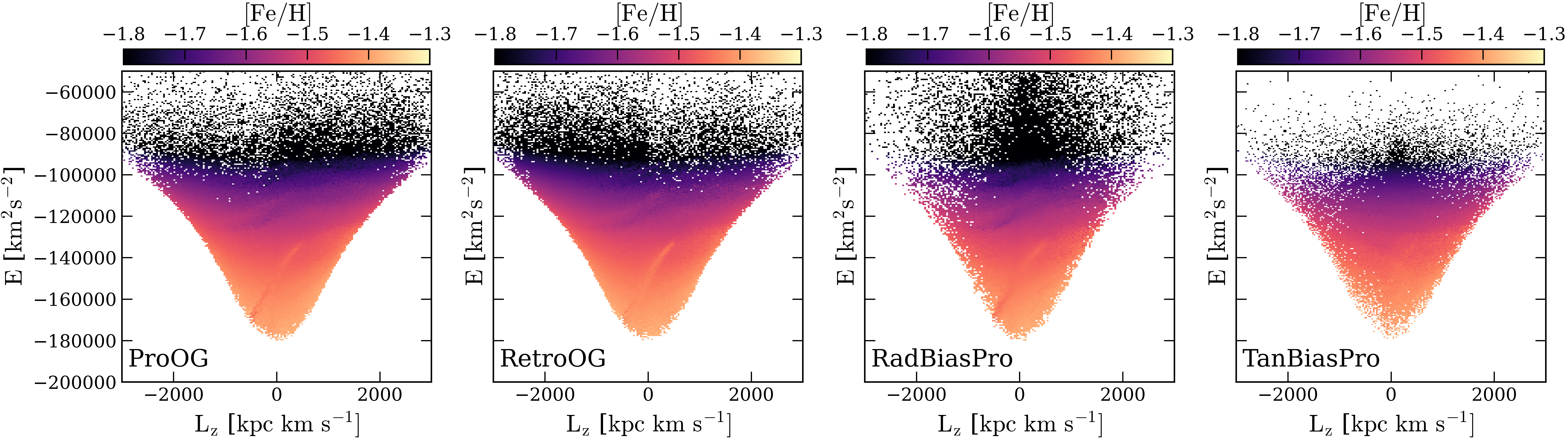}
        \label{fig:ELz_feh_solar}
    \end{subfigure}

    \hfill
    \begin{subfigure}{\linewidth}
        \centering
        \includegraphics[width=\linewidth]{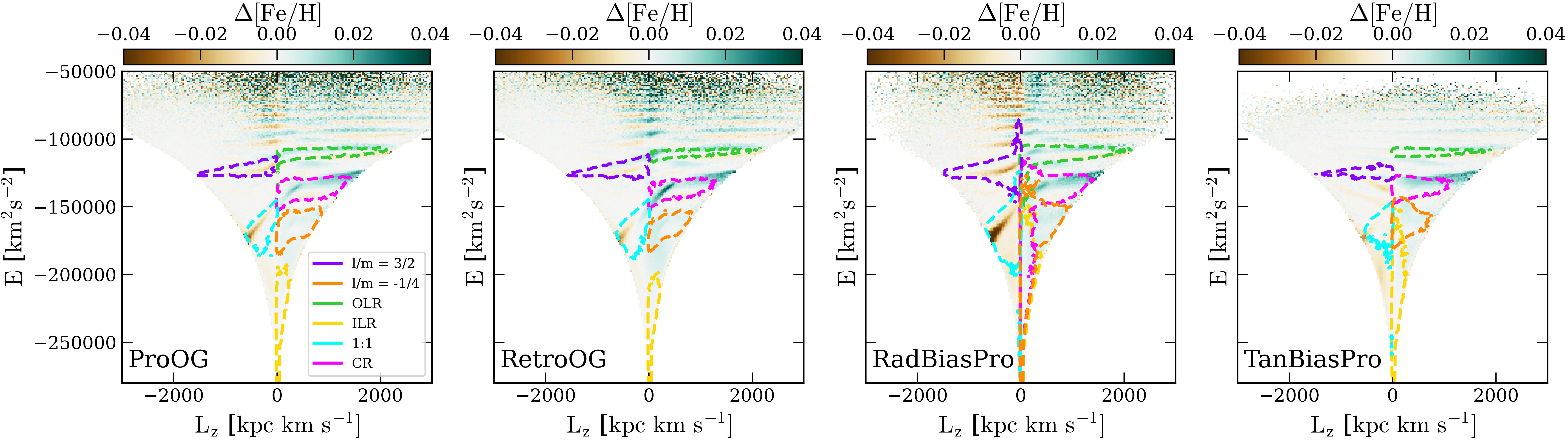}
        \label{fig:ELz_delta_feh}
    \end{subfigure}

        \hfill
    \begin{subfigure}{\linewidth}
        \centering
        \includegraphics[width=\linewidth]{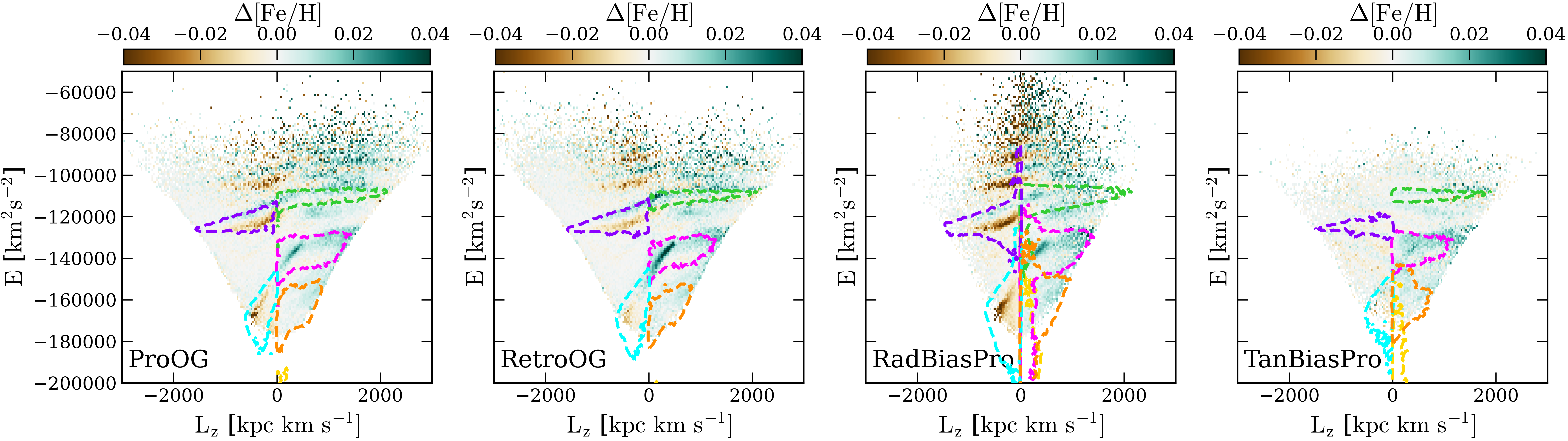}
        \label{fig:ELz_delta_feh_solar}
    \end{subfigure}
    
    \caption{Energy - angular momentum plots at final snapshot showing (a) metallicity distribution of stars within $\mathrm{r \leq 12}$ kpc (b) metallicity distribution of stars within our selection of the 'solar neighbourhood', (c) metallicity contrast between initial and final snapshot $\mathrm{\Delta[Fe/H]=[Fe/H]-[Fe/H]_{0}}$ for all particles within $\mathrm{r \leq 12}$ kpc, and (d) same as (c) but for the 'solar neighbourhood' region. }
    \label{fig:ELz_feh_and_delta_feh}

\end{figure*}

\subsection{Shadow bar} \label{sub:shaddow_bar}

\begin{figure*}
    \centering
    
    \begin{subfigure}{\linewidth}
        \centering
        \includegraphics[width=\linewidth]{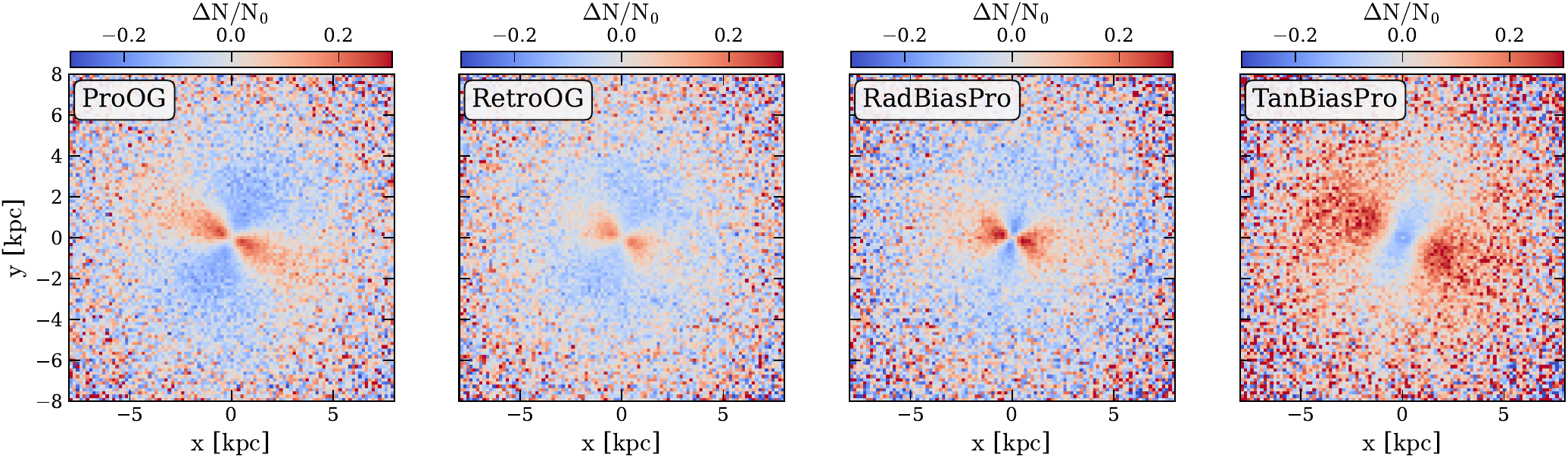}
        \caption{}
        \label{fig:xy_butterfly}
    \end{subfigure}
    \hfill
    \begin{subfigure}{\linewidth}
        \centering
        \includegraphics[width=\linewidth]{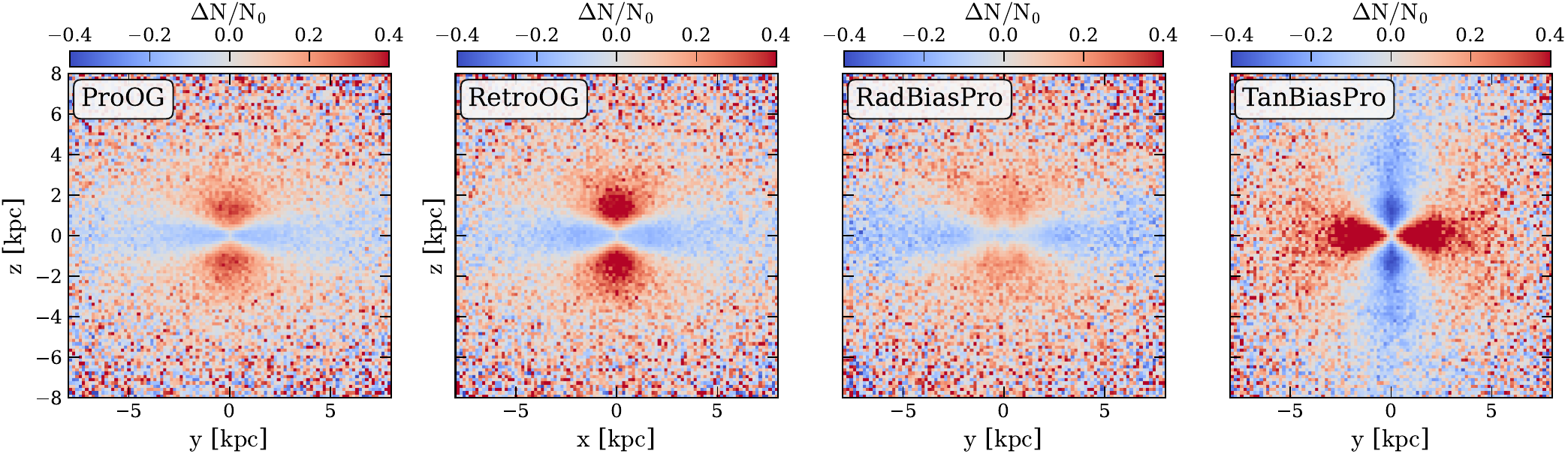}
        \caption{}
        \label{fig:yz_butterfly}
    \end{subfigure}

    \hfill
    \begin{subfigure}{\linewidth}
        \centering
        \includegraphics[width=\linewidth]{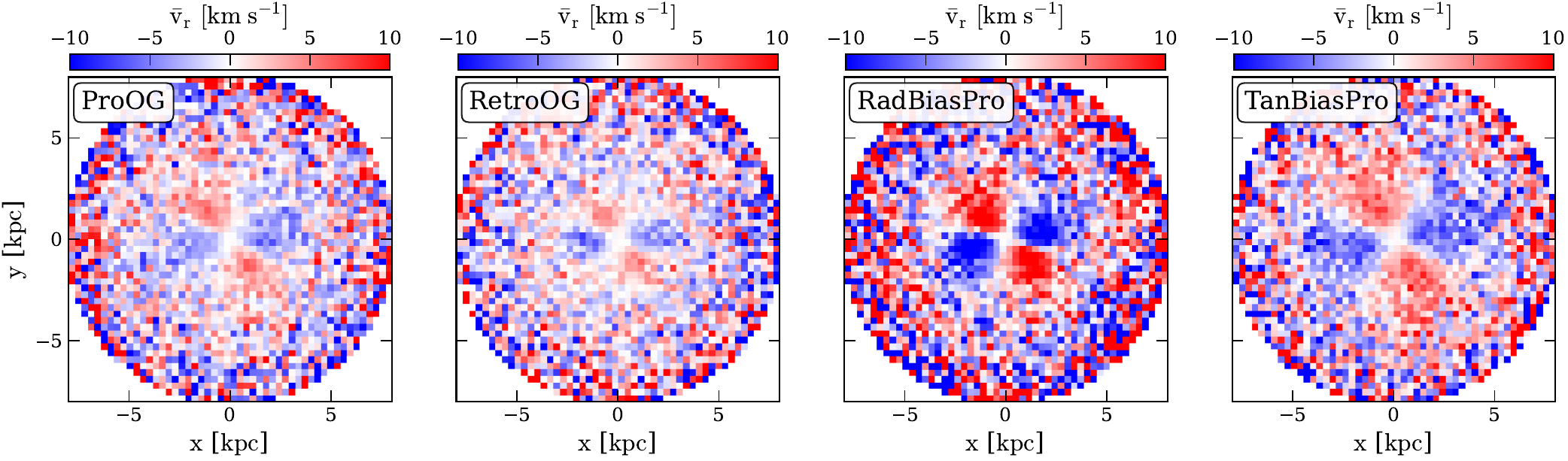}
        \caption{}
        \label{fig:butterfly_vr}
    \end{subfigure}

    \caption{(a) Surface number density contrast in the $x$--$y$ plane, defined as the difference between the final and initial particle distributions, showing a clear quadrupole (``butterfly'') pattern aligned with the bar. (b) The corresponding density contrast in the $x$--$z$ projection, highlighting the vertical redistribution of particles induced by the perturbation. (c) $x$--$y$ projection colour-coded by radial velocity, $v_R$, revealing a quadrupole streaming pattern aligned with the bar. Together, these panels demonstrate the coherent non-axisymmetric response of the halo to the imposed bar potential. }
    \label{fig:shaddow_bar}
    
\end{figure*}

\noindent The spatial response of the halo to the bar perturbation can be seen in Fig.~\ref{fig:xy_butterfly}, where we computed the number density difference between the final and initial particle distribution in $x$-$y$ projection. The resulting density contrast exhibits a characteristic quadrupole, or ``butterfly'', pattern, with overdensities along the bar major axis and underdensities deficits along the minor axis. This morphology is indicative of a bar-like response in the halo population. Although the halo is dynamically hot and largely pressure-supported, the perturbation from the rotating bar is sufficient to induce a coherent redistribution of particles that produces a  bar-like density feature \citep[eg.,][]{Fragkoudi+2017,Debattista+2017}.

\par A similar phenomenon has previously been identified in simulations of DM halos, where the stellar bar induces a bar-shaped overdensity in the surrounding halo through resonant orbital trapping. This response is often referred to as a \textit{shadow bar} \citep[e.g.][]{Petersen+2016_shadow_bar,Hunt+2026_shadow_bar}, as it arises from the dynamical response of halo particles rather than from a self-gravitating instability \citep[eg.,][]{Ostriker+1973}. While most previous studies have focused on DM halos, the same dynamical mechanism is expected to operate for stellar halo populations.

\par Figure~\ref{fig:butterfly_vr} shows the $x$-$y$ projection of the particles colour-coded by their radial velocity, $v_R$, revealing a clear quadrupole streaming pattern aligned with the bar, consistent with the expected kinematic signature of a bar-like substructure. Taken together with the density structure, this demonstrates that the imposed bar potential induces a coherent non-axisymmetric bar-like response in the halo.

\par The properties of the shadow bar vary between the simulations, despite the identical global potential and imposed bar perturbation. In particular, both the spatial extent and amplitude of the quadrupole density signal differ across the models. This dependence arises from the underlying halo DF: resonant trapping selects orbits with specific phase-space properties, so the population of trapped particles is determined by the halo DF. A range of resonant orbit families can contribute to supporting the bar \citep[e.g.,][]{Contopoulos_Grosbol1989}, with the dominant contribution likely arising from x1-type orbit families and higher-order resonances such as the 4:1 ultraharmonic resonance. 

In radially biased halos, the shadow bar is strong but spatially compact. Eccentric orbits are inefficiently trapped because their large radial excursions prevent sustained phase-locking with the bar, confining the response to the inner halo. In contrast, more isotropic and tangentially biased halos contain a larger fraction of near-circular orbits, which are efficiently trapped and remain phase-coherent, producing more extended and strong shadow bars. Intermediate cases therefore yield longer but weaker structures. Furthermore, halos with net prograde rotation further increase the radial extent of the shadow bar.

\section{Resonant dynamics and transport} \label{sec:discussion}

The dynamical response of the halo to the galactic bar can be understood within the framework of resonant dynamics and perturbation theory. In action–angle space, the motion near a resonance can be reduced to that of a simple pendulum through a canonical transformation to the so called 'slow and fast variables' \citep[e.g.][]{Binney2018,Tremaine_Weinberg1984,Chiba_Schonrich2022}.  In this picture, one angle (the resonant angle) and its corresponding action evolves slowly, while the remaining action-angles average out. This leads to libration around the resonance and the formation of resonant islands.

\par We outline the slow–fast action–angle formalism below and refer the interested reader to Appendix \ref{appendix:slow_fast_actions} for the full derivations.

\par A canonical transformation near a resonance $\bm{n}=(n_r,n_\phi)=(l,m)$ is performed to the slow-fast angle action variables $(\bm{\theta',J'})=(\theta_{f},\theta_s,J_{f},J_s)$:
\begin{equation}
    \theta_s=l\theta_r+m(\theta_{\phi}-\Omega_{bar}t), \; \theta_f=\theta_r,
    \label{eq:slow_fast_angle}
\end{equation}

\noindent where $\theta_s$ and $\theta_f$ are the slow and fast angles. The $\theta_s$ angle varies much more slowly (on timescales of $\approx (\bm{n\cdot \Omega}-m\Omega_{bar})^{-1}$) near a resonance compared to $\theta_f$. It can be shown that averaging over the fast motion in the transformed coordinates then reduces the Hamiltonian to that of a classical pendulum::
\begin{equation}
    \bar{H}'(\theta_s,\bm{J}') = H_0(\bm{J}') - m\Omega_{\mathrm{bar}}J_s + 2\left|\Psi_1  \right|\cos\theta_s,
\end{equation}
where $\Psi_1$ is the first-order resonant Fourier component of the perturbing bar potential in the new action variables after averaging over the fast angle.

\begin{figure}
    \centering
    \includegraphics[width=0.8\linewidth]{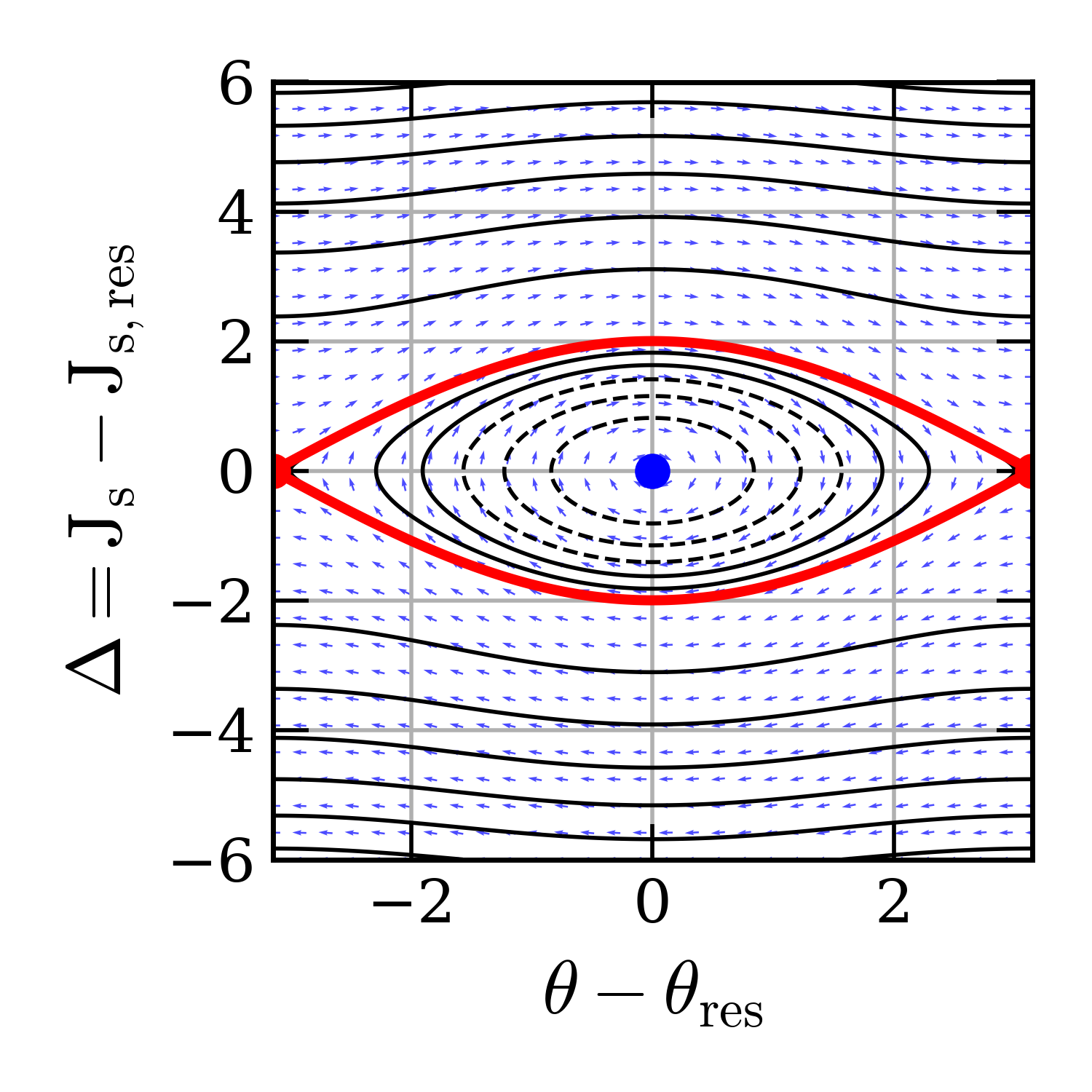}
    \caption{An example of the phase space portrait of a generic simple pendulum.}
    \label{fig:phase_portrait_simple_pendulum}
\end{figure}

Figure \ref{fig:phase_portrait_simple_pendulum} shows an example of a phase portrait for a simple pendulum. The central fixed point (shown in blue) marks orbits that satisfy the resonance condition exactly. The surrounding libration island contains trapped orbits that slowly librate about this point, while the separatrix (red contour) defines the boundary between trapped and circulating motion. The width and location of these islands are set by the bar strength, pattern speed, and the underlying potential. The width of the resonant island is given by:
\begin{equation}
    \Delta J_{\mathrm{res}} = 4\sqrt{\frac{2\left|\Psi_1 \right|}{\partial^2 H_0/\partial J_s^2}}.
\end{equation}

Stars within these resonant islands can become trapped and undergo a secular exchange of angular momentum with the bar. The global response of the halo is set by how these regions are populated, which is encoded by the DF which controls the number of participating particles, while its gradients across the resonance transport direction govern the resulting structures. This is explored further in the next subsection. 

\subsection*{Over- or under-densities?}
Figure~\ref{fig:ELz_summary_allsims} shows that the same resonance can give rise to either over- or underdensities in $E\text{--}L_{z}$ space, as is particularly evident for the retrograde 1:1 resonance. These features are ultimately projections of structures that form in action space as well. The bar induces transport along resonant directions in action space, described by the vector $\bm{\Delta J}_{\rm res} = (\Delta J_r, \Delta L_z)$. \textit{Whether this transport produces an overdensity or an underdensity depends on the gradient of the distribution function $f(\bm{J})$ along this direction}: the response is determined by whether particles are transported from regions of higher phase-space density to lower density, or vice versa.
\par To estimate how the distribution function changes along the direction of resonant transport, we expand the DF to first order around the initial location in action space, $\bm{J}_{0}=(J_{r0},L_{z0},J_{z0})$, of the particle before the perturbation was induced:

\begin{align}
f(J_r, L_z,J_z=J_{z0}) &\approx f(J_{r0}, L_{z0},J_{z0}) + \notag \\
&+ \left.\frac{\partial f}{\partial J_r}\right|_{(J_{r0}, L_{z0})} \Delta J_r \notag
+ \left.\frac{\partial f}{\partial L_z}\right|_{(J_{r0}, L_{z0})} \Delta L_z +  \notag\\ 
&+ \mathcal{O}(\Delta J^2), \notag \\
\Rightarrow \Delta f & \approx \Delta \bm{J}_{\rm res} \cdot \nabla_{\bm{J}} f . \label{eq:delta_df}
\end{align}

\begin{figure*}
    \centering
    
        \begin{subfigure}{\linewidth}
    \centering
    \includegraphics[width=\linewidth]{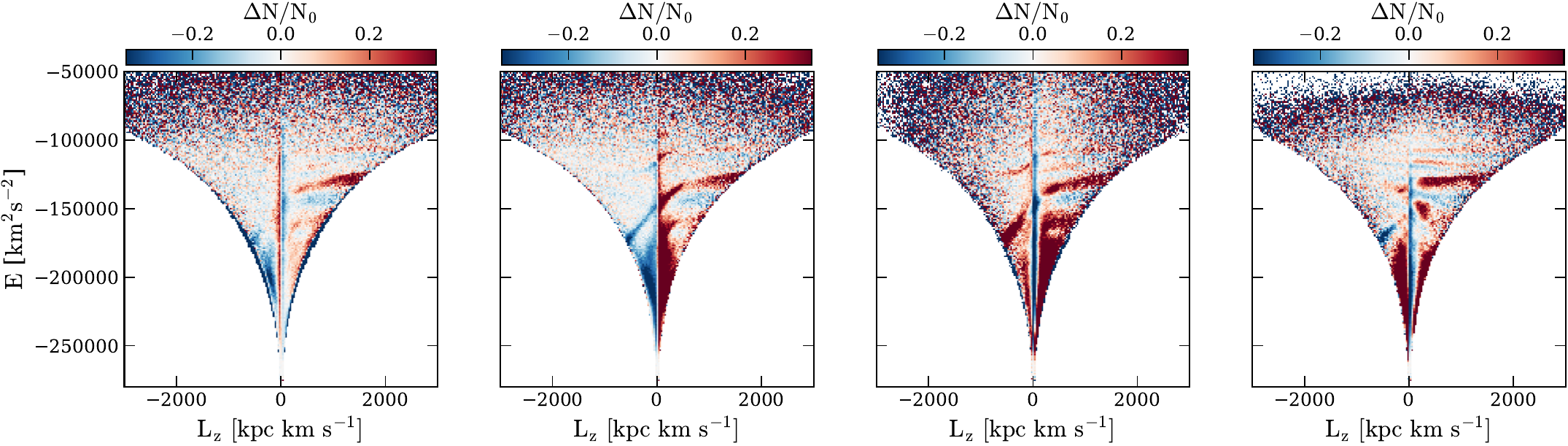}

    \label{fig:over_under_predictions_deltaf}
    \end{subfigure}

    \hfill

    \begin{subfigure}{\linewidth}
    \centering
    \includegraphics[width=\linewidth]{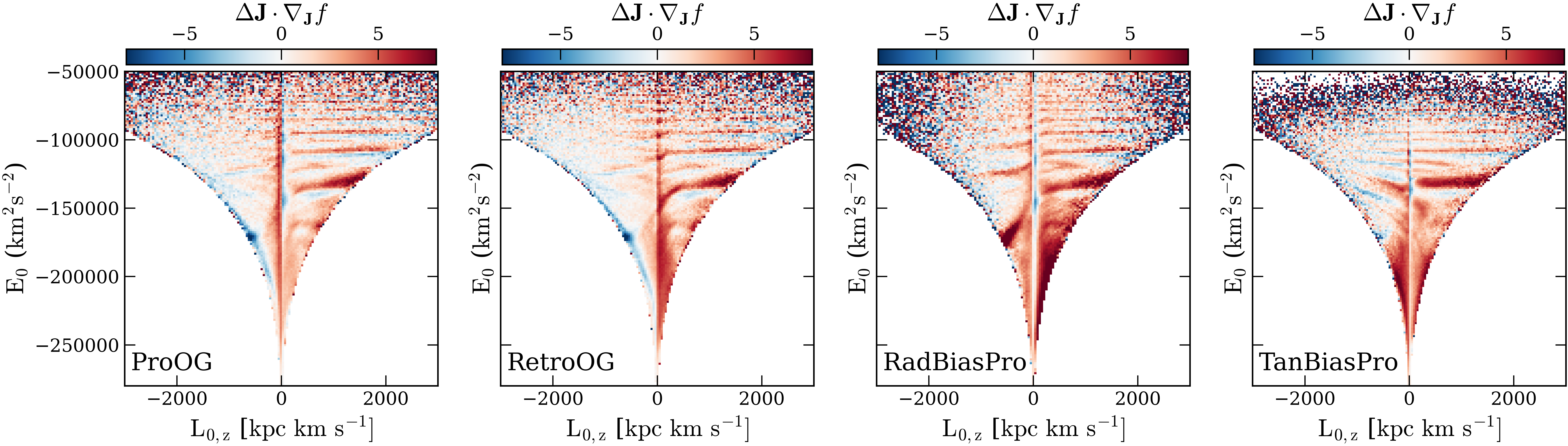}

    \label{fig:over_under_predictions_deltaf}
    \end{subfigure}

    \caption{(a) Final $E-Lz$ distribution of particles in all four simulations the colorbar indicating number density contrast between initial and final snapshots. (b) Initial $E-Lz$ distribution of particles in all four simulations before the bar perturbation is introduced, with the colorbar indicating the change in the phase space density (in action space) along the resonant directions $\Delta f  \approx \Delta \bm{J}_{\rm res} \cdot \nabla_{\bm{J}} f$.}
    \label{fig:ELz_df_predictions}
    
\end{figure*}

In particular it's the sign of $\Delta f$ that indicates if an overdensity ($\Delta f>0$) or underdensity ($\Delta f<0$) is expected to form. In the second row of Fig. \ref{fig:ELz_df_predictions}, we show the initial $E_0-L_{z,0}$ distribution of particles (before the bar perturbation is introduced) with the colour bar indicating the average value of $\Delta f(\bm{J})$ in each bin, computed using Eq.~\ref{eq:delta_df}. The predicted features reproduce well those seen in Fig.~\ref{fig:ELz_summary_allsims} (after the simulation is ran) in terms of their location, geometry and in whether they correspond to over- or under-densities. Of course, the exact relative amplitudes (especially for the higher order resonances) of the substructures are not exactly recovered, which is expected since for this calculation we neglected the evolution of the DF as the (finite number of) particles in our simulation are redistributed in phase space, which would have the effect of making features stronger.

\subsection*{Angular momentum: gained or lost?}

In an otherwise axisymmetric potential perturbed by a rotating bar,
\begin{equation}
    H(\bm{J},\bm{\theta},t) = H_0(\bm{J}) + \sum_{\bm{n}} \Psi_{\bm{n}}(\bm{J}) e^{i(\bm{n}\cdot\bm{\theta} - m\Omega_{\rm bar} t)},
\end{equation}
the torque on an individual orbit is given by
\begin{equation}
    \dot{L}_z = -\frac{\partial H}{\partial \theta_\phi}. \label{eq:dLz_dt}
\end{equation}

Away from resonance, the phase oscillates rapidly and the torque averages to zero over time. Near resonance, however, the phase evolves slowly and remains coherent for long times, allowing small torques to accumulate into a secular change in angular momentum.

\par Individual stars at the same resonance can experience either positive or negative torques depending on their phase relative to the bar. To determine whether a region of phase space undergoes a net gain or loss of angular momentum, we compute the DF-weighted average angular momentum change in that region,
\begin{equation}
\langle \Delta L_z \rangle =
\int \Delta L_z(\bm{J},\bm{\theta})\, f(\bm{J})\,
d^3\bm{J}\, d^3\bm{\theta},
\end{equation}
following the approach of \cite{LBK_1972} (hereafter LBK). This averaging incorporates the contributions of all stars within the relevant phase-space volume and isolates the secular transport induced by the bar, rather than the instantaneous, phase-dependent torques acting on individual orbits. Applying this procedure yields the LBK result:
\begin{align}     
   \langle \dot{\Delta} L_{z} \rangle & = -\frac{1}{8\pi}\iint_{0}^{\infty} m\left(l\frac{\partial f}{\partial J_R} + m\frac{\partial f}{\partial L_z}\right) |\psi_{lm}|^2 \times    \label{eq:LBK} \\ 
   & \times \delta(l\Omega_R+m\Omega_\phi-m\Omega_{bar}) \mathrm{d}J_R \mathrm{d}L_{z}, \notag
\end{align}

which gives the total torque at a resonance $\bm{n}=(l,m)$. If we want to focus on a specific region of phase-space, we change the integration limits of eq. \ref{eq:LBK} accordingly.

It can be seen that the torque $\langle \dot{\Delta} L_{z} \rangle$ depends explicitly on the gradients of the distribution function along the resonant direction in action space, as well as on the local amplitude of the perturbing potential.

\par Similarly, we can apply the same reasoning to the radial action $J_r$:
\begin{equation}
    \dot{J}_r = -\frac{\partial H}{\partial \theta_r}. \label{eq:dJr_dt}
\end{equation}

Integrating both eq. \ref{eq:dLz_dt} and eq. \ref{eq:dJr_dt} in time, it can be shown that the transport in action space along a given resonance occurs along directions given by:
\begin{equation}
    \frac{\Delta J_r}{\Delta L_z} \simeq \frac{l}{m}. \label{eq:delta_Jr_delta_Lz_discussion}
\end{equation}
\vspace{+0.5cm}
\par
\begin{figure*}
    \centering
    \includegraphics[width=\textwidth]{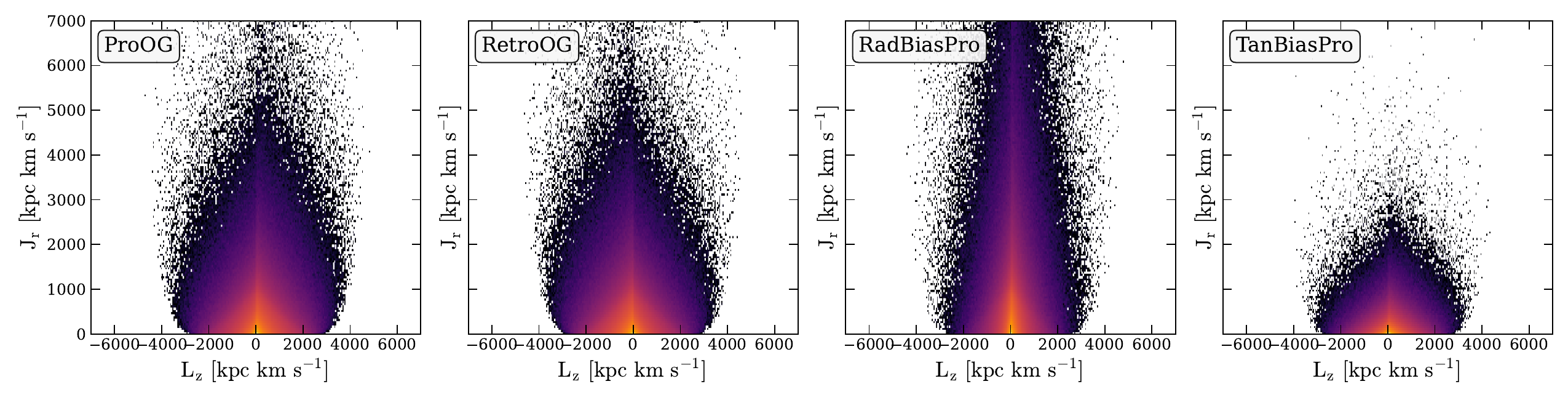}
    \caption{Initial distribution in the $L_z-J_r$ plane of the axisymmetric action space of all four simulations.}
    \label{fig:LzJr_initial_allsims}
\end{figure*}

 The sign of $\langle \dot{\Delta} L_{z} \rangle$ in Eq.~\ref{eq:LBK} determines whether angular momentum is gained or lost at a given resonance. For example, at corotation ($l=0$), the torque sign depends on the azimuthal gradient of the distribution function, i.e. $\partial f/\partial L_z$. For a typical galactic halo distribution function, this gradient is negative for $L_z>0$ (see Fig.~\ref{fig:LzJr_initial_allsims}), reflecting the decreasing phase-space density with increasing angular momentum. As a result, the LBK torque at corotation is positive, implying that stars at this resonance gain angular momentum on average. This is clearly seen in the second row of Fig.~\ref{fig:ELz_final_cbar_delta_actions}, which shows the $E$--$L_z$ distribution of stars colour-coded by their change in angular momentum, $\Delta L_z$.

\par For the retrograde 1:1 resonance, particles in action space evolve approximately along trajectories satisfying $\Delta J_r / \Delta L_z \approx 1$, as given by Eq.~\ref{eq:delta_Jr_delta_Lz_discussion}. From Fig.~\ref{fig:LzJr_initial_allsims} it can be seen that on the retrograde side, the DF gradients satisfy $\partial f/\partial J_r < 0$ and $\partial f/\partial L_z > 0$, while typically the magnitude of the radial gradient is smaller, $|\partial f/\partial J_r| < |\partial f/\partial L_z|$. Therefore, following the LBK torque (Eq.~\ref{eq:LBK}) along the resonant direction yields a net contribution dominated by the $L_z$ gradient, implying $\langle \dot{\Delta} L_z \rangle < 0$. This corresponds to an average loss of angular momentum for stars participating in the retrograde 1:1 resonance (see also \citealt{Tomlinson2026} for a similar phenomenon observed in the stellar halo of a high-resolution cosmological simulation).

\par Also note how for the radially biased halo (RadBiasPro) the fractional change in azimuthal angular momentum $\Delta Lz/L_{z0}$ at the 1:1 resonance is much larger than for the other less radially biased simulations. This is because in the radially biased model, the DF gradients along $J_r$ are shallower compared to the other simulations, while stronger gradients remain along $L_z$, as seen in Fig.~\ref{fig:LzJr_initial_allsims}. This will shift the balance of terms in Eq.~\ref{eq:LBK}, leading to larger angular momentum exchange at this resonance.

\section{Conclusion} \label{sec:conclusion}

In this work, we have explored how dispersion-dominated (i.e. hot pressure supported systems) respond to a rotating bar, focusing on the role of the underlying distribution function of the halo and the orbital anisotropy $\beta$ in particular. Using test-particle simulations in a Milky Way–like potential with a time-dependent bar, we explored how the underlying orbital anisotropy, as well as net prograde or retrograde rotation of halo-like components, shape the bar–halo interaction.

\par The same resonance can produce qualitatively different features in $E-L_z$ space, particularly on the retrograde side, where the 1:1 resonance can generate either overdense or underdense ridges depending on how the resonant islands are populated along the direction of resonant transport induced by the bar. Other main resonances, including corotation, ILR, the $l/m=-1/4$ ultraharmonic, and $l/m=3/2$, also induce features whose morphology and strength vary with the underlying DF. 

In terms of angular momentum transfer, the orbital anisotropy anisotropy plays a key role in setting the strength of resonant torques on the halo. While on average the main prograde resonances (corotation, ILR, and higher-order ultraharmonics) are associated with net angular momentum gain ($\Delta Lz>0$), the retrograde 1:1 resonance dominates the angular momentum exchange on the retrograde side with angular momentum loss on average ($\Delta Lz <0$). In particular, the more radially biased a halo, the stronger the retrograde torque is is at the 1:1 resonance. This is due to the larger imbalance of DF gradients across the resonant transport direction in action space, where $\left| \partial f/\partial J_r \right| \ll \left| \partial f/\partial L_z \right|$, as it can also be seen in Fig. \ref{fig:LzJr_initial_allsims}.

\subsection{Implications for stellar halos}
Our results highlight the need for caution when interpreting structures in $E-L_z$ space as tracers of the Milky Way’s assembly history \citep[e.g.][]{DeLeo+2026,Tomlinson2026}, as bar-driven resonant dynamics can generate substructure that could mimic past accretion events. At the same time, other studies have shown that the bar can also act to disperse pre-existing substructures \citep[e.g.][]{Dilamore_Sanders+2025}, further complicating this interpretation. 

\par A potential silver lining comes from the chemical imprint of resonant features, which may help distinguish structures induced by the bar from those originating in accretion events. Depending on the original metallicity gradient, the resonances can cause either metal rich or metal poor substructures compared to the background (see Fig. \ref{fig:ELz_feh_and_delta_feh}, as well as \citealt{Tomlinson2026}). These chemically distinct patterns can therefore provide a complementary diagnostic, which may help distinguish bar-induced structures from genuinely accreted populations.

\subsection{Implications for dark matter halos}
The implications of our results extend naturally to DM halos. Previous studies have already shown that the efficiency of angular momentum transfer between bars and halos depends not only on the halo density profile, but also on its kinematic structure \citep[e.g.,][]{Athanassoula2003}.

\par Halos with different velocity dispersions exhibit different rates of angular momentum exchange with the bar. Since velocity dispersion reflects the underlying orbital composition of the halo, this can be interpreted within the framework developed here: the distribution function determines the population of resonant orbits and the gradients across resonant phase-space regions. Velocity anisotropy and net prograde or retrograde rotation therefore influence how many and how efficiently resonant particles participate in angular momentum exchange. 

\par Finally, we find that the properties of the shadow bars induced vary between the halo models , despite the identical underlying bar perturbation. Strongly radially biased halos (RadBiasPro) produce strong but compact shadow bars, while tangentially biased halos generate similarly strong but considerably more extended structures. Net rotation also affects the response, with prograde halos producing slightly longer and stronger shadow bars than their retrograde counterparts. This highlights that shadow bars cannot be understood solely in terms of the stellar bars inducing them, but must be analysed in the full context of the underlying phase-space structure of the halo.

\vspace{+0.5cm}

We note that the present study is based on test-particle simulations in which the bar potential is imposed and does not respond to the halo. This controlled setup is intentional, as it allows us to isolate the dynamical response of the halo to a fixed, time-dependent perturbation and cleanly identify how different distribution functions shape resonant transport of halo particles. As a result, we do not capture the fully self-consistent exchange of angular momentum that governs bar slowdown and secular evolution in live systems. Extending this framework to fully self-consistent $N$-body simulations to quantify how the halo DF influences the coupled bar–halo evolution will be explored in future work.

\section*{Acknowledgements}

PG and FF acknowledge support by the UKRI Future Leaders Fellowship (grant no. MR/X033740/1). We acknowledge support from the Science and Technology Facilities Council (STFC) [grant number ST/X001075/1]. We thank Eugene Vasiliev for helpful discussions and for valuable guidance in setting up the initial simulations within the \texttt{Agama} framework.

\section*{Data Availability}
The Agama software package is publicly available and includes example scripts for constructing models similar to those used in this work, as well as routines for computing orbital actions.



\bibliographystyle{mnras}
\bibliography{biblio} 

@ARTICLE{Vasiliev2019_agama,
       author = {{Vasiliev}, Eugene},
        title = "{AGAMA: action-based galaxy modelling architecture}",
      journal = {\mnras},
         year = 2019,
        month = jan,
       volume = {482},
       number = {2},
        pages = {1525-1544},
          doi = {10.1093/mnras/sty2672},
archivePrefix = {arXiv},
       eprint = {1802.08239},
 primaryClass = {astro-ph.GA},
       adsurl = {https://ui.adsabs.harvard.edu/abs/2019MNRAS.482.1525V}
}

@ARTICLE{Sormani+2022,
       author = {{Sormani}, Mattia C. and {Gerhard}, Ortwin and {Portail}, Matthieu and {Vasiliev}, Eugene and {Clarke}, Jonathan},
        title = "{The stellar mass distribution of the Milky Way's bar: an analytical model}",
      journal = {\mnras},
         year = 2022,
        month = jul,
       volume = {514},
       number = {1},
        pages = {L1-L5},
          doi = {10.1093/mnrasl/slac046},
archivePrefix = {arXiv},
       eprint = {2204.13114},
 primaryClass = {astro-ph.GA},
       adsurl = {https://ui.adsabs.harvard.edu/abs/2022MNRAS.514L...1S}
}

@ARTICLE{Hunter+2024,
       author = {{Hunter}, Glen H. and {Sormani}, Mattia C. and {Beckmann}, Jan P. and {Vasiliev}, Eugene and {Glover}, Simon C.~O. and {Klessen}, Ralf S. and {Soler}, Juan D. and {Brucy}, No{\'e} and {Girichidis}, Philipp and {G{\"o}ller}, Junia and {Ohlin}, Loke and {Tress}, Robin and {Molinari}, Sergio and {Gerhard}, Ortwin and {Benedettini}, Milena and {Smith}, Rowan and {Hennebelle}, Patrick and {Testi}, Leonardo},
        title = "{Testing kinematic distances under a realistic Galactic potential: Investigating systematic errors in the kinematic distance method arising from a non-axisymmetric potential}",
      journal = {\aap},
         year = 2024,
        month = dec,
       volume = {692},
          eid = {A216},
        pages = {A216},
          doi = {10.1051/0004-6361/202450000},
archivePrefix = {arXiv},
       eprint = {2403.18000},
 primaryClass = {astro-ph.GA},
       adsurl = {https://ui.adsabs.harvard.edu/abs/2024A&A...692A.216H}
}

@ARTICLE{Portail+2017,
       author = {{Portail}, Matthieu and {Gerhard}, Ortwin and {Wegg}, Christopher and {Ness}, Melissa},
        title = "{Dynamical modelling of the galactic bulge and bar: the Milky Way's pattern speed, stellar and dark matter mass distribution}",
      journal = {\mnras},
         year = 2017,
        month = feb,
       volume = {465},
       number = {2},
        pages = {1621-1644},
          doi = {10.1093/mnras/stw2819},
archivePrefix = {arXiv},
       eprint = {1608.07954},
 primaryClass = {astro-ph.GA},
       adsurl = {https://ui.adsabs.harvard.edu/abs/2017MNRAS.465.1621P}
}

@ARTICLE{Posti+2015,
       author = {{Posti}, Lorenzo and {Binney}, James and {Nipoti}, Carlo and {Ciotti}, Luca},
        title = "{Action-based distribution functions for spheroidal galaxy components}",
      journal = {\mnras},
         year = 2015,
        month = mar,
       volume = {447},
       number = {4},
        pages = {3060-3068},
          doi = {10.1093/mnras/stu2608},
archivePrefix = {arXiv},
       eprint = {1411.7897},
 primaryClass = {astro-ph.GA},
       adsurl = {https://ui.adsabs.harvard.edu/abs/2015MNRAS.447.3060P}
}

@ARTICLE{Tremain_Weinberg1984,
       author = {{Tremaine}, S. and {Weinberg}, M.~D.},
        title = "{A kinematic method for measuring the pattern speed of barred galaxies.}",
      journal = {\apjl},
         year = 1984,
        month = jul,
       volume = {282},
        pages = {L5-L7},
          doi = {10.1086/184292},
       adsurl = {https://ui.adsabs.harvard.edu/abs/1984ApJ...282L...5T}
}

@ARTICLE{Chiba_Schonrich2022,
       author = {{Chiba}, Rimpei and {Sch{\"o}nrich}, Ralph},
        title = "{Oscillating dynamical friction on galactic bars by trapped dark matter}",
      journal = {\mnras},
         year = 2022,
        month = jun,
       volume = {513},
       number = {1},
        pages = {768-787},
          doi = {10.1093/mnras/stac697},
archivePrefix = {arXiv},
       eprint = {2109.10910},
 primaryClass = {astro-ph.GA},
       adsurl = {https://ui.adsabs.harvard.edu/abs/2022MNRAS.513..768C}
}

@ARTICLE{Amorisco2017,
       author = {{Amorisco}, N.~C.},
        title = "{Contributions to the accreted stellar halo: an atlas of stellar deposition}",
      journal = {\mnras},
         year = 2017,
        month = jan,
       volume = {464},
       number = {3},
        pages = {2882-2895},
          doi = {10.1093/mnras/stw2229},
archivePrefix = {arXiv},
       eprint = {1511.08806},
 primaryClass = {astro-ph.GA},
       adsurl = {https://ui.adsabs.harvard.edu/abs/2017MNRAS.464.2882A}
}

@ARTICLE{Trick+2021,
       author = {{Trick}, Wilma H. and {Fragkoudi}, Francesca and {Hunt}, Jason A.~S. and {Mackereth}, J. Ted and {White}, Simon D.~M.},
        title = "{Identifying resonances of the Galactic bar in Gaia DR2: I. Clues from action space}",
      journal = {\mnras},
         year = 2021,
        month = jan,
       volume = {500},
       number = {2},
        pages = {2645-2665},
          doi = {10.1093/mnras/staa3317},
archivePrefix = {arXiv},
       eprint = {1906.04786},
 primaryClass = {astro-ph.GA},
       adsurl = {https://ui.adsabs.harvard.edu/abs/2021MNRAS.500.2645T}
}

@ARTICLE{LBK_1972,
       author = {{Lynden-Bell}, D. and {Kalnajs}, A.~J.},
        title = "{On the generating mechanism of spiral structure}",
      journal = {\mnras},
         year = 1972,
        month = jan,
       volume = {157},
        pages = {1},
          doi = {10.1093/mnras/157.1.1},
       adsurl = {https://ui.adsabs.harvard.edu/abs/1972MNRAS.157....1L}
}

@ARTICLE{Tremaine_Weinberg1984,
       author = {{Tremaine}, S. and {Weinberg}, M.~D.},
        title = "{Dynamical friction in spherical systems.}",
      journal = {\mnras},
         year = 1984,
        month = aug,
       volume = {209},
        pages = {729-757},
          doi = {10.1093/mnras/209.4.729},
       adsurl = {https://ui.adsabs.harvard.edu/abs/1984MNRAS.209..729T}
}

@ARTICLE{Binney2018,
       author = {{Binney}, James},
        title = "{Orbital tori for non-axisymmetric galaxies}",
      journal = {\mnras},
         year = 2018,
        month = feb,
       volume = {474},
       number = {2},
        pages = {2706-2724},
          doi = {10.1093/mnras/stx2835},
archivePrefix = {arXiv},
       eprint = {1710.11360},
 primaryClass = {astro-ph.GA},
       adsurl = {https://ui.adsabs.harvard.edu/abs/2018MNRAS.474.2706B}
}

@ARTICLE{Petersen+2016_shadow_bar,
       author = {{Petersen}, Michael S. and {Weinberg}, Martin D. and {Katz}, Neal},
        title = "{Dark matter trapping by stellar bars: the shadow bar}",
      journal = {\mnras},
         year = 2016,
        month = dec,
       volume = {463},
       number = {2},
        pages = {1952-1967},
          doi = {10.1093/mnras/stw2141},
archivePrefix = {arXiv},
       eprint = {1602.04826},
 primaryClass = {astro-ph.GA},
       adsurl = {https://ui.adsabs.harvard.edu/abs/2016MNRAS.463.1952P}
}

@ARTICLE{Hunt+2026_shadow_bar,
       author = {{Hunt}, Jason A.~S. and {Petersen}, Michael S. and {Weinberg}, Martin D. and {Johnston}, Kathryn V. and {Bernet}, Marcel and {Daniel}, Kathryne J. and {Hyman}, S{\'o}ley {\'O}. and {Price-Whelan}, Adrian M. and {Arora}, Arpit},
        title = "{The dark matter wake of a galactic bar revealed by multichannel singular spectral analysis}",
      journal = {\mnras},
         year = 2026,
        month = jan,
       volume = {545},
       number = {3},
          eid = {staf2118},
        pages = {staf2118},
          doi = {10.1093/mnras/staf2118},
archivePrefix = {arXiv},
       eprint = {2510.09751},
 primaryClass = {astro-ph.GA},
       adsurl = {https://ui.adsabs.harvard.edu/abs/2026MNRAS.545f2118H}
}

@ARTICLE{Weinberg1985,
       author = {{Weinberg}, M.~D.},
        title = "{Evolution of barred galaxies by dynamical friction.}",
      journal = {\mnras},
         year = 1985,
        month = mar,
       volume = {213},
        pages = {451-471},
          doi = {10.1093/mnras/213.3.451},
       adsurl = {https://ui.adsabs.harvard.edu/abs/1985MNRAS.213..451W}
}

@ARTICLE{Athanassoula2003,
       author = {{Athanassoula}, E.},
        title = "{What determines the strength and the slowdown rate of bars?}",
      journal = {\mnras},
         year = 2003,
        month = jun,
       volume = {341},
       number = {4},
        pages = {1179-1198},
          doi = {10.1046/j.1365-8711.2003.06473.x},
archivePrefix = {arXiv},
       eprint = {astro-ph/0302519},
 primaryClass = {astro-ph},
       adsurl = {https://ui.adsabs.harvard.edu/abs/2003MNRAS.341.1179A}
}

@ARTICLE{Anatoja2018,
       author = {{Antoja}, T. and {Helmi}, A. and {Romero-G{\'o}mez}, M. and {Katz}, D. and {Babusiaux}, C. and {Drimmel}, R. and {Evans}, D.~W. and {Figueras}, F. and {Poggio}, E. and {Reyl{\'e}}, C. and {Robin}, A.~C. and {Seabroke}, G. and {Soubiran}, C.},
        title = "{A dynamically young and perturbed Milky Way disk}",
      journal = {\nat},
         year = 2018,
        month = sep,
       volume = {561},
       number = {7723},
        pages = {360-362},
          doi = {10.1038/s41586-018-0510-7},
archivePrefix = {arXiv},
       eprint = {1804.10196},
 primaryClass = {astro-ph.GA},
       adsurl = {https://ui.adsabs.harvard.edu/abs/2018Natur.561..360A}
}

@ARTICLE{Fragkoudi2019,
       author = {{Fragkoudi}, F. and {Katz}, D. and {Trick}, W. and {White}, S.~D.~M. and {Di Matteo}, P. and {Sormani}, M.~C. and {Khoperskov}, S. and {Haywood}, M. and {Hall{\'e}}, A. and {G{\'o}mez}, A.},
        title = "{On the ridges, undulations, and streams in Gaia DR2: linking the topography of phase space to the orbital structure of an N-body bar}",
      journal = {\mnras},
         year = 2019,
        month = sep,
       volume = {488},
       number = {3},
        pages = {3324-3339},
          doi = {10.1093/mnras/stz1875},
archivePrefix = {arXiv},
       eprint = {1901.07568},
 primaryClass = {astro-ph.GA},
       adsurl = {https://ui.adsabs.harvard.edu/abs/2019MNRAS.488.3324F}
}

@ARTICLE{Weinberg_Katz2007,
       author = {{Weinberg}, Martin D. and {Katz}, Neal},
        title = "{The bar-halo interaction - I. From fundamental dynamics to revised N-body requirements}",
      journal = {\mnras},
         year = 2007,
        month = feb,
       volume = {375},
       number = {2},
        pages = {425-459},
          doi = {10.1111/j.1365-2966.2006.11306.x},
archivePrefix = {arXiv},
       eprint = {astro-ph/0508166},
 primaryClass = {astro-ph},
       adsurl = {https://ui.adsabs.harvard.edu/abs/2007MNRAS.375..425W}
}

@ARTICLE{Collier_Madigan2021,
       author = {{Collier}, Angela and {Madigan}, Ann-Marie},
        title = "{The Coupling of Galactic Dark Matter Halos with Stellar Bars}",
      journal = {\apj},
         year = 2021,
        month = jul,
       volume = {915},
       number = {1},
          eid = {23},
        pages = {23},
          doi = {10.3847/1538-4357/ac004d},
archivePrefix = {arXiv},
       eprint = {2105.04698},
 primaryClass = {astro-ph.GA},
       adsurl = {https://ui.adsabs.harvard.edu/abs/2021ApJ...915...23C}
}

@ARTICLE{Dilamore+2023,
       author = {{Dillamore}, Adam M. and {Belokurov}, Vasily and {Evans}, N. Wyn and {Davies}, Elliot Y.},
        title = "{Stellar halo substructure generated by bar resonances}",
      journal = {\mnras},
         year = 2023,
        month = sep,
       volume = {524},
       number = {3},
        pages = {3596-3608},
          doi = {10.1093/mnras/stad2136},
archivePrefix = {arXiv},
       eprint = {2303.00008},
 primaryClass = {astro-ph.GA},
       adsurl = {https://ui.adsabs.harvard.edu/abs/2023MNRAS.524.3596D}
}

@ARTICLE{Tomlinson2026,
       author = {{Tomlinson}, Thomas and {Fragkoudi}, Francesca and {Carrillo}, Andreia and {Fattahi}, Azadeh and {Gherghinescu}, Paula and {Deason}, Alis and {Pakmor}, R{\"u}diger and {Grand}, Robert J.~J. and {G{\'o}mez}, Facundo A. and {van de Voort}, Freeke and {Bieri}, Rebekka},
        title = "{Stirring Things Up: Bar-induced substructures in the stellar halo of a cosmological Milky Way analogue}",
      journal = {arXiv e-prints},
         year = 2026,
        month = jan,
          eid = {arXiv:2601.14409},
        pages = {arXiv:2601.14409},
          doi = {10.48550/arXiv.2601.14409},
archivePrefix = {arXiv},
       eprint = {2601.14409},
 primaryClass = {astro-ph.GA},
       adsurl = {https://ui.adsabs.harvard.edu/abs/2026arXiv260114409T}
}

@ARTICLE{Chiba+2021,
       author = {{Chiba}, Rimpei and {Friske}, Jennifer K.~S. and {Sch{\"o}nrich}, Ralph},
        title = "{Resonance sweeping by a decelerating Galactic bar}",
      journal = {\mnras},
         year = 2021,
        month = jan,
       volume = {500},
       number = {4},
        pages = {4710-4729},
          doi = {10.1093/mnras/staa3585},
archivePrefix = {arXiv},
       eprint = {1912.04304},
 primaryClass = {astro-ph.GA},
       adsurl = {https://ui.adsabs.harvard.edu/abs/2021MNRAS.500.4710C}
}

@ARTICLE{Dilamore_Sanders+2025,
       author = {{Dillamore}, Adam M. and {Sanders}, Jason L.},
        title = "{Bar-driven dispersal of Galactic substructure}",
      journal = {\mnras},
         year = 2025,
        month = sep,
       volume = {542},
       number = {2},
        pages = {1331-1346},
          doi = {10.1093/mnras/staf1264},
archivePrefix = {arXiv},
       eprint = {2506.09117},
 primaryClass = {astro-ph.GA},
       adsurl = {https://ui.adsabs.harvard.edu/abs/2025MNRAS.542.1331D}
}

@ARTICLE{Debattista_Sellwood2000,
       author = {{Debattista}, Victor P. and {Sellwood}, J.~A.},
        title = "{Constraints from Dynamical Friction on the Dark Matter Content of Barred Galaxies}",
      journal = {\apj},
         year = 2000,
        month = nov,
       volume = {543},
       number = {2},
        pages = {704-721},
          doi = {10.1086/317148},
archivePrefix = {arXiv},
       eprint = {astro-ph/0006275},
 primaryClass = {astro-ph},
       adsurl = {https://ui.adsabs.harvard.edu/abs/2000ApJ...543..704D}
}

@ARTICLE{Fragkoudi+2015,
       author = {{Fragkoudi}, F. and {Athanassoula}, E. and {Bosma}, A. and {Iannuzzi}, F.},
        title = "{The effects of Boxy/Peanut bulges on galaxy models}",
      journal = {\mnras},
         year = 2015,
        month = jun,
       volume = {450},
       number = {1},
        pages = {229-245},
          doi = {10.1093/mnras/stv537},
archivePrefix = {arXiv},
       eprint = {1503.03068},
 primaryClass = {astro-ph.GA},
       adsurl = {https://ui.adsabs.harvard.edu/abs/2015MNRAS.450..229F}
}

@ARTICLE{Fragkoudi+2016,
       author = {{Fragkoudi}, F. and {Athanassoula}, E. and {Bosma}, A.},
        title = "{A close look at secular evolution: boxy/peanut bulges reduce gas inflow to the central kiloparsec}",
      journal = {\mnras},
         year = 2016,
        month = oct,
       volume = {462},
       number = {1},
        pages = {L41-L45},
          doi = {10.1093/mnrasl/slw120},
archivePrefix = {arXiv},
       eprint = {1606.04540},
 primaryClass = {astro-ph.GA},
       adsurl = {https://ui.adsabs.harvard.edu/abs/2016MNRAS.462L..41F}
}

@ARTICLE{Hunt+2019,
       author = {{Hunt}, Jason A.~S. and {Bub}, Mathew W. and {Bovy}, Jo and {Mackereth}, J. Ted and {Trick}, Wilma H. and {Kawata}, Daisuke},
        title = "{Signatures of resonance and phase mixing in the Galactic disc}",
      journal = {\mnras},
         year = 2019,
        month = nov,
       volume = {490},
       number = {1},
        pages = {1026-1043},
          doi = {10.1093/mnras/stz2667},
archivePrefix = {arXiv},
       eprint = {1904.10968},
 primaryClass = {astro-ph.GA},
       adsurl = {https://ui.adsabs.harvard.edu/abs/2019MNRAS.490.1026H}
}

@ARTICLE{Kawata+2018,
       author = {{Kawata}, Daisuke and {Baba}, Junichi and {Ciuc{\v{a}}}, Ioana and {Cropper}, Mark and {Grand}, Robert J.~J. and {Hunt}, Jason A.~S. and {Seabroke}, George},
        title = "{Radial distribution of stellar motions in Gaia DR2}",
      journal = {\mnras},
         year = 2018,
        month = sep,
       volume = {479},
       number = {1},
        pages = {L108-L112},
          doi = {10.1093/mnrasl/sly107},
archivePrefix = {arXiv},
       eprint = {1804.10175},
 primaryClass = {astro-ph.GA},
       adsurl = {https://ui.adsabs.harvard.edu/abs/2018MNRAS.479L.108K}
}

@ARTICLE{Dehnen+2000,
       author = {{Dehnen}, Walter},
        title = "{The Effect of the Outer Lindblad Resonance of the Galactic Bar on the Local Stellar Velocity Distribution}",
      journal = {\aj},
         year = 2000,
        month = feb,
       volume = {119},
       number = {2},
        pages = {800-812},
          doi = {10.1086/301226},
archivePrefix = {arXiv},
       eprint = {astro-ph/9911161},
 primaryClass = {astro-ph},
       adsurl = {https://ui.adsabs.harvard.edu/abs/2000AJ....119..800D}
}

@ARTICLE{Fux+2001,
       author = {{Fux}, R.},
        title = "{Order and chaos in the local disc stellar kinematics induced by the Galactic bar}",
      journal = {\aap},
         year = 2001,
        month = jul,
       volume = {373},
        pages = {511-535},
          doi = {10.1051/0004-6361:20010561},
archivePrefix = {arXiv},
       eprint = {astro-ph/0105398},
 primaryClass = {astro-ph},
       adsurl = {https://ui.adsabs.harvard.edu/abs/2001A&A...373..511F}
}

@ARTICLE{Fragkoudi+2021,
       author = {{Fragkoudi}, F. and {Grand}, R.~J.~J. and {Pakmor}, R. and {Springel}, V. and {White}, S.~D.~M. and {Marinacci}, F. and {Gomez}, F.~A. and {Navarro}, J.~F.},
        title = "{Revisiting the tension between fast bars and the {\ensuremath{\Lambda}}CDM paradigm}",
      journal = {\aap},
         year = 2021,
        month = jun,
       volume = {650},
          eid = {L16},
        pages = {L16},
          doi = {10.1051/0004-6361/202140320},
archivePrefix = {arXiv},
       eprint = {2011.13942},
 primaryClass = {astro-ph.GA},
       adsurl = {https://ui.adsabs.harvard.edu/abs/2021A&A...650L..16F}
}

@ARTICLE{Cooper+2010,
       author = {{Cooper}, A.~P. and {Cole}, S. and {Frenk}, C.~S. and {White}, S.~D.~M. and {Helly}, J. and {Benson}, A.~J. and {De Lucia}, G. and {Helmi}, A. and {Jenkins}, A. and {Navarro}, J.~F. and {Springel}, V. and {Wang}, J.},
        title = "{Galactic stellar haloes in the CDM model}",
      journal = {\mnras},
         year = 2010,
        month = aug,
       volume = {406},
       number = {2},
        pages = {744-766},
          doi = {10.1111/j.1365-2966.2010.16740.x},
archivePrefix = {arXiv},
       eprint = {0910.3211},
 primaryClass = {astro-ph.GA},
       adsurl = {https://ui.adsabs.harvard.edu/abs/2010MNRAS.406..744C}
}

@ARTICLE{Bullock_Johnston2005,
       author = {{Bullock}, James S. and {Johnston}, Kathryn V.},
        title = "{Tracing Galaxy Formation with Stellar Halos. I. Methods}",
      journal = {\apj},
         year = 2005,
        month = dec,
       volume = {635},
       number = {2},
        pages = {931-949},
          doi = {10.1086/497422},
archivePrefix = {arXiv},
       eprint = {astro-ph/0506467},
 primaryClass = {astro-ph},
       adsurl = {https://ui.adsabs.harvard.edu/abs/2005ApJ...635..931B}
}

@ARTICLE{He+2024,
       author = {{He}, Jiaxin and {Wang}, Wenting and {Li}, Zhaozhou and {Han}, Jiaxin and {Rodriguez-Gomez}, Vicente and {Zhao}, Donghai and {Meng}, Xianguang and {Jing}, Yipeng and {Shao}, Shi and {Shi}, Rui and {Tan}, Zhenlin},
        title = "{How Do the Velocity Anisotropies of Halo Stars, Dark Matter, and Satellite Galaxies Depend on Host Halo Properties?}",
      journal = {\apj},
         year = 2024,
        month = dec,
       volume = {976},
       number = {2},
          eid = {187},
        pages = {187},
          doi = {10.3847/1538-4357/ad8882},
archivePrefix = {arXiv},
       eprint = {2407.14827},
 primaryClass = {astro-ph.GA},
       adsurl = {https://ui.adsabs.harvard.edu/abs/2024ApJ...976..187H}
}

@ARTICLE{Zhang+2026,
       author = {{Zhang}, Xiuyuan and {Thoyas}, Andreas and {Necib}, Lina and {Wetzel}, Andrew and {Arora}, Arpit},
        title = "{Set the Night on FIRE: Building an Empirical Local Dark Matter Velocity Distribution}",
      journal = {arXiv e-prints},
         year = 2026,
        month = mar,
          eid = {arXiv:2603.25783},
        pages = {arXiv:2603.25783},
          doi = {10.48550/arXiv.2603.25783},
archivePrefix = {arXiv},
       eprint = {2603.25783},
 primaryClass = {astro-ph.GA},
       adsurl = {https://ui.adsabs.harvard.edu/abs/2026arXiv260325783Z}
}

@ARTICLE{Dillamore+2024,
       author = {{Dillamore}, Adam M. and {Belokurov}, Vasily and {Evans}, N. Wyn},
        title = "{Radial halo substructure in harmony with the Galactic bar}",
      journal = {\mnras},
         year = 2024,
        month = aug,
       volume = {532},
       number = {4},
        pages = {4389-4407},
          doi = {10.1093/mnras/stae1789},
archivePrefix = {arXiv},
       eprint = {2402.14907},
 primaryClass = {astro-ph.GA},
       adsurl = {https://ui.adsabs.harvard.edu/abs/2024MNRAS.532.4389D}
}

@BOOK{Binney_Tremaine_2008,
       author = {{Binney}, James and {Tremaine}, Scott},
        title = "{Galactic Dynamics: Second Edition}",
         year = 2008,
       adsurl = {https://ui.adsabs.harvard.edu/abs/2008gady.book.....B}
}

@article{Cole_Lacey1996,
    author = {Cole, Shaun and Lacey, Cedric},
    title = {The structure of dark matter haloes in hierarchical clustering models},
    journal = {Monthly Notices of the Royal Astronomical Society},
    volume = {281},
    number = {2},
    pages = {716-736},
    year = {1996},
    month = {07},
    issn = {0035-8711},
    doi = {10.1093/mnras/281.2.716},
    url = {https://doi.org/10.1093/mnras/281.2.716},
    eprint = {https://academic.oup.com/mnras/article-pdf/281/2/716/3204113/281-2-716.pdf},
}

@ARTICLE{Hansen_Moore2006,
       author = {{Hansen}, Steen H. and {Moore}, Ben},
        title = "{A universal density slope   Velocity anisotropy relation for relaxed structures}",
      journal = {\na},
         year = 2006,
        month = mar,
       volume = {11},
       number = {5},
        pages = {333-338},
          doi = {10.1016/j.newast.2005.09.001},
archivePrefix = {arXiv},
       eprint = {astro-ph/0411473},
 primaryClass = {astro-ph},
       adsurl = {https://ui.adsabs.harvard.edu/abs/2006NewA...11..333H}
}

@ARTICLE{Sellwood+2002,
       author = {{Sellwood}, J.~A. and {Binney}, J.~J.},
        title = "{Radial mixing in galactic discs}",
      journal = {\mnras},
         year = 2002,
        month = nov,
       volume = {336},
       number = {3},
        pages = {785-796},
          doi = {10.1046/j.1365-8711.2002.05806.x},
archivePrefix = {arXiv},
       eprint = {astro-ph/0203510},
 primaryClass = {astro-ph},
       adsurl = {https://ui.adsabs.harvard.edu/abs/2002MNRAS.336..785S}
}

@ARTICLE{Debattista+2017,
       author = {{Debattista}, Victor P. and {Ness}, Melissa and {Gonzalez}, Oscar A. and {Freeman}, K. and {Zoccali}, Manuela and {Minniti}, Dante},
        title = "{Separation of stellar populations by an evolving bar: implications for the bulge of the Milky Way}",
      journal = {\mnras},
         year = 2017,
        month = aug,
       volume = {469},
       number = {2},
        pages = {1587-1611},
          doi = {10.1093/mnras/stx947},
archivePrefix = {arXiv},
       eprint = {1611.09023},
 primaryClass = {astro-ph.GA},
       adsurl = {https://ui.adsabs.harvard.edu/abs/2017MNRAS.469.1587D}
}

@ARTICLE{Fragkoudi+2017,
       author = {{Fragkoudi}, F. and {Di Matteo}, P. and {Haywood}, M. and {G{\'o}mez}, A. and {Combes}, F. and {Katz}, D. and {Semelin}, B.},
        title = "{Bars and boxy/peanut bulges in thin and thick discs. I. Morphology and line-of-sight velocities of a fiducial model}",
      journal = {\aap},
         year = 2017,
        month = oct,
       volume = {606},
          eid = {A47},
        pages = {A47},
          doi = {10.1051/0004-6361/201630244},
archivePrefix = {arXiv},
       eprint = {1704.00734},
 primaryClass = {astro-ph.GA},
       adsurl = {https://ui.adsabs.harvard.edu/abs/2017A&A...606A..47F}
}

@ARTICLE{Contopoulos_Grosbol1989,
       author = {{Contopoulos}, G. and {Grosbol}, P.},
        title = "{Orbits in barred galaxies}",
      journal = {\aapr},
         year = 1989,
        month = nov,
       volume = {1},
       number = {3-4},
        pages = {261-289},
          doi = {10.1007/BF00873080},
       adsurl = {https://ui.adsabs.harvard.edu/abs/1989A&ARv...1..261C}
}

@ARTICLE{DeLeo+2026,
       author = {{De Leo}, M. and {Massari}, D. and {Bellazzini}, M. and {Mucciarelli}, A. and {Acosta-Tripailao}, B. and {Nipoti}, C.},
        title = "{Dynamical mirages: How bar-induced resonant trapping can mimic substructure clustering in dynamical parameter spaces}",
      journal = {\aap},
         year = 2026,
        month = mar,
       volume = {707},
          eid = {A310},
        pages = {A310},
          doi = {10.1051/0004-6361/202556723},
archivePrefix = {arXiv},
       eprint = {2511.05655},
 primaryClass = {astro-ph.GA},
       adsurl = {https://ui.adsabs.harvard.edu/abs/2026A&A...707A.310D}
}

@ARTICLE{Ostriker+1973,
       author = {{Ostriker}, J.~P. and {Peebles}, P.~J.~E.},
        title = "{A Numerical Study of the Stability of Flattened Galaxies: or, can Cold Galaxies Survive?}",
      journal = {\apj},
         year = 1973,
        month = dec,
       volume = {186},
        pages = {467-480},
          doi = {10.1086/152513},
       adsurl = {https://ui.adsabs.harvard.edu/abs/1973ApJ...186..467O}
}




\appendix
\section{Slow and fast action-angles formalism} \label{appendix:slow_fast_actions}
We can reduce the dynamical evolution near a resonance to that of a simple pendulum through a canonical transformation to the \textit{slow and fast action-angle variables} \citep[e.g.][]{Binney2018,Tremaine_Weinberg1984,Chiba_Schonrich2022}. In this picture, the resonant degree of freedom evolves slowly, and the remaining action–angle variables vary rapidly and are averaged over. The dynamics then reduces to libration islands about the resonance (as can be seen in Fig. \ref{fig:phase_portrait_simple_pendulum}).

\par We outline the slow–fast action–angle formalism below for completeness. The discussion follows closely that of found in Appendix B of \cite{Chiba+2021}.

\par We perform a canonical transformation near a resonance $\bm{n}=(n_r,n_\phi)=(l,m)$ to the slow-fast angle action variables $(\bm{\theta',J'})=(\theta_{f},\theta_s,J_{f},J_s)$:
\begin{equation}
    \theta_s=l\theta_r+m(\theta_{\phi}-\Omega_{bar}t), \; \theta_f=\theta_r,
    \label{eq:slow_fast_angle_appendix}
\end{equation}

\noindent where are the slow and fast angles respectively. The $\theta_s$ angle varies much more slowly (on timescales of $\approx (\bm{n\cdot \Omega}-m\Omega_{bar})^{-1}$) near a resonance compared to $\theta_f$. To get the new action variables $\bm{J'}$ we perform a canonical transformation to the new coordinates using a generating function $W(\bm{\theta},\bm{J'},t)$, such that
\begin{align}
    W(\bm{\theta},\bm{J'},t) &= [l\theta_r+m(\theta_\phi-\Omega_{bar}t)]J_s+\theta_r J_f, \\
    \bm{\theta'} &= \frac{\partial W}{\partial \bm{J'}}, \; \bm{J}=\frac{\partial W}{\partial \bm{\theta}}, \\
    H'(\bm{\theta',J'},t) &= H(\bm{\theta,J},t) + \frac{\partial W}{\partial t}.
    \label{eq:transformation_slow_fast_appendix}
\end{align}
The Hamiltonian in the new coordinate system then becomes:
\begin{align}
  H'(\bm{\theta'}, \bm{J'}) &= H_0(\bm{J'}) - m \Omega_\mathrm{bar} J_s 
    + \sum_{\bm{k}} \Psi_{\bm{k}} e^{i \bm{k} \cdot \bm{\theta'}}, \\
  \bm{k} &= (k_f, k_s),
\end{align}
where $H_0$ is the axisymmetric, unperturbed Hamiltonian before the bar perturbation is induced. Near resonance, the fast angle $\theta_f$ evolves much faster than the slow one $\theta_s$, so we can average the Hamiltonian $H'$ over the fast motion which is not of interest:
\begin{equation}
    \bar{H}'(\theta_s,J_s) = \frac{1}{2\pi}\int \mathrm{d}\theta_f H'(\bm{\theta',J'}),
\end{equation}
such that:
\begin{equation}
    \bar{H}'(\theta_s,J_s) = H_0(\bm{J'}) - m\Omega_{bar}J_s + \sum_{k_s}\Psi_{k_s}e^{ik_s\theta_s}.
\end{equation}
For major resonances for which $m=2$ it can be shown that only $\Psi_{k_s}$ is non-zero only at $k_s=\pm 1$ \citep[e.g. Appendix B][]{Chiba+2021}, reducing the Hamiltonian to that of the classic pendulum:
\begin{equation}
    \bar{H}'(\theta_s,\bm{J}') = H_0(\bm{J}') - m\Omega_{\mathrm{bar}}J_s + 2\left|\Psi_1  \right|\cos\theta_s.
\end{equation}


\bsp	
\label{lastpage}
\end{document}